\documentclass[a4paper, 12pt]{article}
\usepackage[english]{babel}
\usepackage[paper=portrait,pagesize]{typearea}
\usepackage{amsmath}
\usepackage{ amssymb }
\usepackage{graphicx}
\usepackage{subcaption}
\usepackage[title]{appendix}
\usepackage[parfill]{parskip}
\usepackage[margin=2cm]{geometry}
\usepackage{amssymb}
\usepackage[normalem]{ulem}
\newcommand{\Hlule}{\rule{\linewidth}{0.5mm}}
\usepackage{bm}

\usepackage{fancyhdr}
\usepackage{enumitem}
\usepackage{natbib}
\usepackage{siunitx}
\usepackage{color,soul}

\usepackage{subfiles}
\usepackage{arydshln}
\usepackage{pdflscape}
\usepackage{afterpage}
\usepackage{rotating}
\usepackage{adjustbox}
\usepackage{apalike}
\usepackage{rotating}
\usepackage{floatrow}
\usepackage{multirow}
\usepackage{url}
\usepackage{comment}
\excludecomment{mysection}
\usepackage{booktabs, tabularx}
\newcolumntype{C}{>{\centering\arraybackslash}X}

\newcommand{\Ml}{} 

\makeatletter
\def\Ml{%
  \@ifnextchar^%
    {\@Ml}
    {
    \@Ml^{}}%
}
\def\@Ml^#1{%
  \mathbf{M}^{l #1}%
}
\makeatother

\newcommand{\Mls}{} 

\makeatletter
\def\Mls{%
  \@ifnextchar^%
    {\@Mls}
    {
    \@Mls^{}}%
}
\def\@Mls^#1{%
  \mathbf{M}^{ls #1}%
}
\makeatother

\newcommand{\Msl}{} 

\makeatletter
\def\Msl{%
  \@ifnextchar^%
    {\@Msl}
    {
    \@Msl^{}}%
}
\def\@Msl^#1{%
  \mathbf{M}^{sl #1}%
}
\makeatother

\newcommand{\Ms}{} 

\makeatletter
\def\Ms{%
  \@ifnextchar^%
    {\@Ms}
    {
    \@Ms^{}}%
}
\def\@Ms^#1{%
  \mathbf{M}^{s #1}%
}
\makeatother

\renewcommand{\subsectionmark}[1]{}
\numberwithin{equation}{section}

\setcitestyle{authoryear,open={(},close={)}}

\usepackage[dvipsnames]{xcolor}
\usepackage[normalem]{ulem}

\usepackage{cancel}
\usepackage{capt-of}

\definecolor{mygreen}{rgb}{0.0, 0.0, 0.0}
\definecolor{myblue}{rgb}{0.0, 0.0, 0.0}

\title{TITLE}
\author{Zackary Bell} 
\date{\today}

\begin{document}
\begin{titlepage}
\begin{center}


\Hlule \\[0.4cm] 
{\large \bfseries Exploring the characteristics of a vehicle-based temperature dataset for convection-permitting numerical weather prediction \par}\vspace{0.4cm} 
\Hlule \\[1.0cm] 

\begin{minipage}[t]{\textwidth}
\large
Authors:
Zackary Bell$^{1*}$, Sarah L. Dance$^{1,2}$ and Joanne A. Waller$^{3}$ %
\begin{center}
$^{1}$Department of Meteorology, University of Reading, Reading, UK \\
$^{2}$Department of Mathematics and Statistics, University of Reading, Reading, UK \\
$^{3}$Met Office@Reading, University of Reading, Reading, UK \\
\end{center}
\end{minipage}\\[0.5cm]

\uline{\large{Abstract:}}

\begin{minipage}{\textwidth}
\qquad Crowdsourced vehicle-based observations have the potential to improve forecast skill in convection-permitting numerical weather prediction (NWP). The aim of this paper is to explore the characteristics of vehicle-based observations of air temperature. We describe a novel low-precision vehicle-based observation dataset obtained from a Met Office proof-of-concept trial. In this trial, observations of air temperature were obtained from built-in vehicle air-temperature sensors, broadcast to an application on the participant's smartphone and uploaded, with relevant metadata, to the Met Office servers. We discuss the instrument and representation uncertainties associated with vehicle-based observations and present a new quality-control procedure. It is shown that, for some observations, location metadata may be inaccurate due to unsuitable smartphone application settings. The characteristics of the data that passed quality-control are examined through comparison with United Kingdom variable-resolution model data, roadside weather information station observations, and Met Office integrated data archive system observations. Our results show that the uncertainty associated with vehicle-based observation-minus-model comparisons is likely to be weather-dependent and possibly vehicle-dependent. Despite the low precision of the data, vehicle-based observations of air temperature could be a useful source of spatially-dense and temporally-frequent observations for NWP. 

\vspace{3mm}


\vspace{1mm}

\textbf{Correspondence}: *Zackary Bell, Department of Meteorology, University of Reading, Earley Gate, Whiteknights Road, Reading, RG6 6ET, United Kingdom, z.n.bell@pgr.reading.ac.uk


\end{minipage}

\vfill
\end{center}
\end{titlepage}

\section{Introduction} \label{sec: introduction}

Convection-permitting numerical weather prediction (NWP) requires a large number of observations of high spatio-temporal resolution to constrain short-term forecasts \citep{sun2014use,gustafsson2018survey,dance2019improvements}. However, due to the cost of installation, management, and maintenance of observing instrumentation, it may be impractical to extend traditional scientific observing networks to provide sufficient additional relevant observations. A potential alternative source of inexpensive observations is from opportunistic data generated by the public or other organisations \citep{waller2020editorial,BLAIR2021100156}.

The application of opportunistic datasets in NWP has been a popular area of research in recent years \citep{hintz2019collecting}. Observations from personal weather stations (PWSs) \citep{steeneveld2011quantifying,wolters2012estimating,chapman2017can,meier2017crowdsourcing,nipen2019adopting} and smartphones \citep{overeem2013crowdsourcing,droste2017crowdsourcing,madaus2017evaluating,hintz2019collectingprocessing,hintz2020estimation,hintz2021crowd} are commonly obtained through crowdsourcing. Such observations may be inaccurate when compared with traditional scientific observations. However, the number of crowdsourced observations available has the potential to far exceed the number of scientific surface observations currently produced \citep{muller2015crowdsourcing}. Opportunistic datasets can also be obtained from partnerships with other organisations. For example, roadside weather information station (RWIS) data obtained from highways agencies are currently assimilated into the Met Office United Kingdom variable-resolution (UKV) model \citep{gustafsson2018survey}. 

Observations obtained from vehicles are another dataset of opportunity \citep{mahoney2013realizing}. Similarly to PWS and smartphone observations, vehicle-based observations can be obtained through crowdsourcing and will therefore be most densely distributed in urban areas and on major transport networks. Vehicle-based observations can also be obtained through several non-crowdsourcing methods. For example, the data can be obtained directly from vehicle manufacturers through connected vehicle initiatives \citep[e.g.,][]{mahoney2013realizing}, from built-in sensors of vehicle fleets via the controller area network (CAN) \citep[e.g.,][]{mercelis2020towards}, or through externally mounted sensors \citep[e.g.,][]{anderson2012quality}. In this paper, vehicle-based observations of air temperature are obtained from built-in vehicle sensors through on-board diagnostic (OBD) dongles. This method of data collection, which is described in section \ref{sec: Methodology}, could be used for crowdsourcing vehicle-based observations. 

Vehicle-based observations are currently used to improve road weather modelling \citep[e.g.,][]{hu2019modeling} and forecasts to combat adverse road weather conditions on transportation networks \citep[e.g.,][]{karsisto2016using,siems2019adaptive}. \citet{karsisto2019verification} showed that assimilation of vehicle-based observations into the Finnish Meteorological Institute's road weather model
had the greatest forecast impact factor when RWISs were sparse. The use of vehicle-based observations in NWP is still in its infancy, but their use for nowcasting has been investigated by the German weather service (DWD) \citep{hintz2019collecting}. Additionally, an observing simulation system experiment (OSSE) conducted by \citet{siems2020impacts} showed a modest but appreciable impact from assimilating simulated vehicle-based observations.


Before opportunistic datasets can be assimilated, they must undergo thorough quality-control (QC) and the contributions to their observation uncertainty identified and investigated. \citet{bell2015good} attributed the total uncertainty of crowdsourced PWS observations to five sources; calibration issues, communication and software issues, inaccurate metadata, design flaws, and error due to unresolved scales. As a result of these issues, which also apply to other opportunistic datasets, the implementation of QC procedures can become substantially more difficult than the QC for traditional observations. In some studies over half the crowdsourced data were removed by the QC procedure (e.g. \citet{meier2017crowdsourcing,madaus2017evaluating,hintz2019collectingprocessing}). \citet{siems2019adaptive} developed QC for vehicle-based observations from disparate sources for use in road weather forecasting systems. However, these QC tests required a large number of observations to be in close spatio-temporal proximity such that spatial comparisons be used. Due to a lack of observations occurring at similar locations and times for the dataset examined in this study, a new QC procedure was developed.

Understanding the characteristics of opportunistic observations is key to their effective use in NWP \citep{waller2020editorial}. For data assimilation, an understanding of the instrument and representation errors that contribute to the total observation uncertainty is required. Important meteorological features such as sharp discontinuities caused by precipitation processes can be observed by opportunistic observations but will likely be misrepresented by a NWP model \citep{mahoney2013realizing}. Hence, it is likely that there will be significant representation error caused by the mis-match in scales observed and modelled \citep{janjic2018representation}. The instrument and representation components of the vehicle-based observation uncertainty are discussed in section \ref{sec: Uncertainties in vehicle-based temperature observations}. For the vehicle-based observations of air temperature examined in this study, an important physical feature misrepresented by a NWP model will be the underlying road surface. The influence of roads on the air temperature measured by vehicles will be complex as the road-surface energy balance at a given location is substantially affected by the availability of water, the quantity of visible sky, and the amount of traffic \citep[e.g.,][]{anandakumar1999study,chapman2011spatial,oke2017urban,karsisto2019verification}. To properly understand the discrepancy between what is observed and modelled, it is necessary to examine the characteristics of the differences between the model and the observations. The objective of this paper is to explore the characteristics of a vehicle-based temperature dataset through comparison with other datasets. 

The format of this paper is as follows. In section \ref{sec: Uncertainties in vehicle-based temperature observations} the uncertainties associated with vehicle-based observations of air temperature are discussed. The Met Office trial used to obtain the vehicle-based observations in this study, the datasets used for comparison, and the novel quality-control procedure applied to the vehicle-based observations are detailed in section \ref{sec: Methodology}. The results of the new quality-control process highlight that the observation location metadata can be inaccurate due to poor GPS signal and application settings. A comparison between vehicle-based observations and other datasets is given in section \ref{sec: Examination of quality controlled dataset}. Our novel results show that the uncertainty of vehicle-based observations is likely weather-dependent and possibly vehicle-dependent. In section \ref{sec: conclusion} our results are summarised and we conclude that vehicle-based observations are a promising opportunistic dataset for convection-permitting data assimilation.

\section{Uncertainties in vehicle-based observations of air temperature} \label{sec: Uncertainties in vehicle-based temperature observations}

\subsection{Vehicle-based observations of air temperature from built-in sensors} \label{sec: Vehicle-based temperature measurements from built-in sensors}

Most modern vehicles are equipped with a sensor to measure the air temperature of the surrounding atmosphere. Throughout this paper, these sensors will be referred to as external air-temperature sensors. Measurements obtained from external air-temperature sensors are used by vehicle air conditioning systems to adjust cabin air temperature \citep{abdelhamid2014vehicle} and alert the driver to safety hazards such as the possible presence of ice on the roads \citep{padarthy2019identification}. External air-temperature sensors are commonly negative temperature coefficient thermistors \citep{fierceelectronics2014temperature}. The location of external air-temperature sensors will vary with vehicle make and manufacturer. Common placements are usually in the airflow at the front of the vehicle, such as behind the grill near the bottom of the vehicle or in the wing mirror \citep{tchir2016why}. We note that most vehicles also have a sensor that measures the air temperature inside the vehicle engine commonly referred to as the intake air-temperature sensor. These measurements, however, are contaminated with heat from the vehicle engine and hence will not be representative of the true atmospheric conditions.

\subsection{Instrument error} \label{sec: Instrument errors}

Built-in vehicle sensors are not intended to give high-quality meteorological information. As such, observations of air temperature from external air-temperature sensors are likely to have substantial instrument uncertainty. There are several sources of instrument uncertainty for vehicle-based observations of air temperature:
\begin{enumerate}
    \item 
    The observations may be affected by extraneous influences \citep{mahoney2013realizing}.
    \item
    The sensing instrument may not be as accurate or precise as required for meteorological applications \citep{mahoney2013realizing}.
    \item
    The ventilation of the sensing instrument may be inadequate \citep{harrison2015meteorological}. 
\end{enumerate}
We now discuss these issues in more detail.

The extraneous influences that vehicle-based observations of air temperature are subject to include heating from the vehicle engine or the underlying road surface. The degree of vehicle influence on the observations will be determined by the sensors proximity to the vehicle engine. \citet{mercelis2020towards} found that observations of air temperature from external air-temperature sensors situated far away from the vehicle engine were consistent with reliable observations obtained from road weather information stations. In contrast, observations obtained from sensors near the vehicle engine had to be discarded due to sensor biases. While external air-temperature sensor placement is usually chosen to mitigate the influence of engine heat \citep{tchir2016why}, radiation reflected from the road surface can be incident on the sensor. Observations of air temperature from external air-temperature sensors in such circumstances may be warmer than the true ambient conditions.

The precision of an observation will depend on the number of significant figures available for the digital representation of the measured value. (The concept we have called precision is known in metrology as resolution \citep{vim2012}).  The difference between a continuous variable and its imprecise digital representation is known as the quantization error \citep{widrow1996statistical}. As the sensing instruments used for opportunistic datasets are not intended to give high-quality meteorological information, quantization uncertainty will likely be part of the instrument uncertainty \citep[e.g.,][]{mirza2016comparison}.

Adequate sensor ventilation is necessary to ensure accurate observations of air temperature \citep{harrison2020shall}. Sensor ventilation for external air-temperature sensors is determined by how fast the vehicle is moving \citep[e.g.,][]{knight2010mapping}.

\subsection{Representation error} \label{sec: Representation error}

Representation error is defined as the difference between a perfect observation and a model's representation of that observation \citep{janjic2018representation,bell2020accounting}. The model's representation of an observation is calculated using an observation operator. An observation operator is function that maps the model state into observation space. According to \citet{janjic2018representation}, the representation error consists of three components:
\begin{enumerate}
    \item 
    The pre-processing error caused by the incorrect preparation of an observation.
    \item
    The observation operator error due to any incorrect or approximate observation operators used in the assimilation of an observation.
    \item
    The error due to unresolved scales and processes when there is a mis-match in scales and processes observed and modelled.
\end{enumerate} 
We now discuss these errors in more detail.

The pre-processing error for vehicle-based observations of air temperature can be caused by the data collection and quality-control procedures. The height of external air-temperature sensors will vary with vehicle type and sensor-height metadata will likely be unavailable in the collection of crowdsourced datasets. Hence, the observations must be assigned a height which may differ from the true height resulting in a height assignment error. The quality-control for the vehicle-based observations is discussed in section \ref{sec: Quality control}.

Since air temperature is usually an NWP variable, the observation operator for vehicle-based observations may be a simple interpolation operator. An observation operator error may result from the misrepresentation of the vehicle-based observation height by the NWP model. The resolution of NWP models is likely to be too coarse to represent the elevation of the vehicle-based observations properly \citep{waller2020evaluating}. This mismatch in elevation between a surface observation and a NWP model field is normally accounted for by correcting the observation to be at the same height as the model field. As air temperature is expected to change with altitude in the surface layer \citep{stull1988introduction}, the model height selected by the observation operator will influence the value of the model-equivalent observation.

For vehicle-based observation of air temperature, errors due to unresolved scales and processes are likely to be caused by deficiencies in the modelling of the local road-surface energy balance (RSEB) between the net radiation into the road surface, the heating caused by traffic, the ground heat flux density, the sensible heat flux density and  the latent heat flux density \citep{karsisto2019verification}. The amount of radiation absorbed by the road will vary across the road due to the sky-view factor and  traffic effects \citep{chapman2011spatial}. The sky-view factor indicates the amount of shielding from radiative heating and cooling and may be highly spatially variable due to trees or buildings near the road. \citet{chapman2011spatial} showed a rural example where the sky-view factor caused road-surface temperature to vary by almost $3^\circ$C.  Traffic effects include the generation of turbulence by vehicles, friction heat dissipation from tyres, sensible heat flux from vehicle engines, heat and moisture from exhaust fumes, and the blocking of incoming solar radiation and outgoing longwave radiation from the road surface \citep{prusa2002conceptual,chapman2005influence}. \citet{gustavsson2001road} found that, during morning commuting hours in urban areas, traffic caused the road surface temperature to increase by approximately $2^\circ$C.

The materials used for road surfaces have a large heat capacity such that much of the radiation absorbed by the surface is converted into ground heat flux \citep{anandakumar1999study}. The remaining turbulent heat fluxes are determined by the amount of water available at the road surface \citep{oke2017urban}. If the road surface is dry, the remaining energy is entirely converted to sensible heat, which will result in a strong vertical air-temperature gradient near the road surface.  Conversely, if water is available at the road surface, some of the remaining energy is converted into latent heat and the air-temperature profile near the surface will be more uniform. 

A common approach for modelling surface fluxes in NWP is through tile schemes \citep[e.g.,][]{essery2003explicit}. Using this approach, the surface flux of a grid box is the weighted average of several different surface fluxes and hence may differ from the local RSEB substantially.

For road forecasting applications, outputs from NWP are post-processed in order to take better account of the road physics \citep[e.g.,][]{clark1998,coulson2012}. For this initial study, we do not use these post-processing techniques for simplicity. However, in principle, a more sophisticated observation operator could use a similar approach to road forecasting models and reduce the uncertainty due to unresolved scales.

\section{Methodology} \label{sec: Methodology}

\subsection{The Met Office trial} \label{sec: The Met Office trial}

From 20th February 2018 to 30th April 2018 the Met Office ran a proof-of-concept trial to collect vehicle-based observations of air temperature. The measuring instruments used in this trial are those built-in by the manufacturer of the vehicle. In this trial, on-board diagnostic (OBD) dongles were used to broadcast reports from the vehicle engine management interface to an application (app) installed on a participant's smartphone via Bluetooth. Additional metadata derived from the smartphone was appended to the report, and uploaded to the Met Office Weather Observations Website \citep{kirk2020weather} using the smartphone's connection to the mobile network (3G etc). A complete description of this trial can be found in \citet{bell2020quality}. We now note some of the important aspects of the trial below.

The data collection frequency and GPS update period was set to $1$ minute while the minimum distance for a GPS update was set to $500$ metres. We also note that a known fault that occurred during this trial was for engine-intake temperature (i.e. the air temperature inside the vehicle engine) to be recorded as air temperature for some observations. Observations were collected throughout the United Kingdom with their locations corresponding to journeys undertaken by the participants. 

The dataset obtained through this trial consists of $67959$ reports obtained from 31 Met Office volunteers. Each report contains some combination of observations of air temperature, engine-intake temperature, and air pressure from built-in vehicle sensors. The observations of temperature have a precision of $1^\circ$C while the observations of air pressure have a precision of $10$hPa. We limit the scope of this study to air temperature only as engine-intake temperature will not reflect the true atmospheric-air temperature and air pressure has too low precision to be useful in NWP. The metadata for each report include vehicle speed (km/h), date-time (given by the application as date and 24 hour clock time), GPS location, vehicle ID, and an unique observation ID. With the exception of vehicle speed which was obtained by the OBD dongle, all metadata was derived by the smartphone app.

\subsection{Additional datasets used in this study} \label{sec: Datasets used in this study}

\subsubsection{Met Office Integrated Data Archive System data} \label{sec: Met Office Integrated Data Archive System data}

Met Office Integrated Data Archive System (MIDAS) temperature data consists of observations of $1.25$m-air temperature which have a precision of $0.1^\circ$C and an uncertainty of $0.2^\circ$C for various locations in the UK \citep{met2012met}. We use MIDAS daily maximum and minimum temperature data in our quality-control procedure described in section \ref{sec: Quality control}. We also use MIDAS hourly temperature data to provide a comparison with vehicle-based observations of air temperature that are within $1.5$km of a MIDAS station (see section \ref{sec: Statistical analysis of observation-minus-background departures}). These data are linearly interpolated to the time of a vehicle-based observation.

\subsubsection{NWP model data} \label{sec: NWP model data}

To explore the characteristics of the vehicle-based observations that pass quality-control we use Met Office $10$-minute UK variable-resolution (UKV) model data \citep{met2016ukv}. The UKV is a variable resolution configuration of the Unified Model whose domain covers the United Kingdom and Ireland \citep{lean2008characteristics}. The inner domain has grid boxes of size $1.5\text{km} \times 1.5\text{km}$ and fully covers the United Kingdom \citep{milan2020hourly}. Surrounding this is a variable-resolution grid with boxes whose edges steadily increase in zonal and/or meridional directions to $4$km in size. 

The UKV model fields we use in this study are $1.5$m-air temperature and surface-air temperature defined as the air temperature at the boundary with the surface. The UKV model data are interpolated to the time and horizontal location of a vehicle-based observation so that we can construct two observation-minus-background (OMB) datasets (i.e. one OMB dataset using surface-air data for the background and another OMB dataset using $1.5$m-air temperature for the background). Since a vehicle-based observation and the horizontally interpolated background are both estimates of the true air temperature, their difference is equal to the difference of their errors. If their errors are independent, the variance of their differences will be equal to the sum of their individual error variances. Therefore, examining the statistics of the two observation-minus-background datasets will provide insight into the uncertainty of the vehicle-based observations. As the height of the external air-temperature sensor for each vehicle is unknown, we are unable to interpolate the model data to the height of a vehicle-based observation or correct the vehicle-based observation to be at the height of either UKV model field. It is likely that the vehicle-based observations are between the two model heights and are closer to the surface than the $1.5$m height. 

The surface flux for each grid box is determined by expressing the percentage of land use as a combination of $5$ vegetation and $4$ non-vegetation tiles \citep{essery2003explicit,porson2010implementation}. For each grid box, the surface flux is obtained by calculating the sum of the weighted average of the fluxes from each tile (where instantaneous interaction between tiles is neglected). The UKV uses the urban canopy model MORUSES (Met Office-Reading Urban Surface Exchange Scheme) as the urban tile. MORUSES represents the impervious urban surface through a roof tile and a canyon tile \citep{hertwig2020urban}. However, observations taken on motorways and major routes will often be surrounded by rural areas, and so the road fraction of the UKV grid box will be small. For example, a typical UK motorway traversing a rural grid box occupies less than $2\%$ of the total area \citep{bremner2019ever}.

\subsubsection{Roadside weather information station observations}

Vehicle-based observations of air temperature from built-in sensors are known to be consistent with reliable observations obtained from roadside weather information stations (RWISs), provided the external air-temperature sensor is located away from the engine block \citep{mercelis2020towards}. We therefore use RWIS data provided by \citet{rwis2018he} to provide a comparison with similar point observations for different weather conditions. There are over $250$ RWISs in England located along major roads and major routes providing various roadside meteorological information with a temporal frequency of $10$ minutes \citep{swis20guide}. In this study, we use RWIS observations of air temperature that have precision of at least $0.1^\circ$C. To give an indication of the total uncertainty of these observations, we note that the  Met Office currently assimilate RWIS observations of air temperature into the UKV with an uncertainty of $1^\circ$C. We note that the height that RWISs measure air temperature is estimated to be between $2$ and $3$ metres, but can be outside of this range if the site is located on a bank \citep{rwis2020height}. Road-state classifiers (i.e. dry, trace amounts of water, wet) provided by RWISs are used to indicate the availability of water at the road surface. The RWIS observations are linearly interpolated to the time that a vehicle passed a station.


\subsection{Quality-control} \label{sec: Quality control}

In this section, we briefly describe the quality-control (QC) process applied to the vehicle-based dataset. Further details are given by \citet{bell2020quality}. We note that, due to the size and spatio-temporal sparsity of this dataset, we were unable to use spatial consistency QC tests.

Before the QC process was implemented, an initial filtering of the raw data from the trial was performed to ensure each observation had an air temperature observation and the relevant metadata needed for each test. This filtering removed $35780$ observations due to either a missing air temperature observation or an invalid speed. The resultant dataset will be referred to as the filtered dataset. 

The QC process applied to the vehicle-based dataset began with three tests applied in parallel: the climatological range test (CRT), the stuck instrument test (SIT), and the global positioning system (GPS) test. Lastly, observations that passed each of these tests were then put through a sensor ventilation test (SVT). The final quality-controlled dataset (QC-dataset) consisted of all observations that passed the SVT. We now provide a brief description of each QC test.

The CRT checked if an observation was within a specified tolerance of a location-specific climatology. For this dataset, we used MIDAS daily temperature data \citep{met2012met} to create monthly climatology datasets. These datasets were constructed by determining the maximum and minimum air temperature of each MIDAS station active during February to April 2018 from pre-2018 data. The CRT was implemented by comparing the observation to the nearest (in terms of great circle distance) MIDAS station monthly climatology dataset. If the observation was within a $2^\circ$C tolerance of the climatological range of the MIDAS station, then the observation was passed.

The SIT examined portions of vehicle-specific time-series to check whether the vehicle sensor was stuck on an air temperature value. This test required a vehicle identifier to determine observations that came from the same source. (This may be unavailable in other crowdsourced observation studies due to data privacy concerns). The SIT was implemented by comparing an observation with all other observations from the same vehicle that occurred within a $15$-minute time-window. If there was at least one observation that had a different value of air temperature to the tested observation, then the tested observation was passed. This test is essentially a simplified version of a persistence test (see \citet{zahumensky2004guidelines} for guidelines) that is able to account for any short journeys undertaken by participants during the trial and the low precision of the data.

The GPS test compared the location of an observation, denoted the test observation, relative to a prior observation from the same vehicle, denoted the reference observation, to evaluate the plausibility of the observation location metadata. The reference observation was at most $30$ minutes before the test observation. As with the SIT, a vehicle identifier was required to determine if observations came from the same source. The GPS test was implemented by calculating the great-circle distance between the test and reference observations, $d_{test}$. Then $d_{test}$ was compared with the maximum and minimum distances estimated using the speed and time metadata for the vehicle. The maximum distance was estimated by
\begin{align} \label{eq: maximum estimated distance}
    d_{max}^e = max(v_{test}, v_{ref}) \times \Delta t,
\end{align}
where $v_{test}$ and $v_{ref}$ are the speeds of the test and reference observations respectively, and $\Delta t$ is the time-gap between the two observations. Similarly, the minimum distance was estimated by
\begin{align} \label{eq: minimum estimated distance}
    d_{min}^e = min(v_{test}, v_{ref}) \times \Delta t.
\end{align}
The test observation passed the GPS test provided $\Gamma_{min} d_{min}^e \leq d_{test} \leq \Gamma_{max} d_{max}^e$ where $\Gamma_{min} = 0.6$ and $\Gamma_{max} = 1.3$ are minimum and maximum multiplicative tolerance constants, respectively. For justification of the choice of $\Gamma_{min}$ and $\Gamma_{max}$, we refer the reader to \citet{bell2020quality}. Test observations with $\Delta t < 1$ minute or $max(v_{test}, v_{ref}) < 25$km/h were passed if $d_{test} \leq \Gamma_{max} d_{max}^e$ as they were expected to be close to the reference observation. (The specific choice of $25$km/h is related to the sensor ventilation test discussed in the next paragraph). If a test observation did not have an observation from the same vehicle that occurred at most $30$ minutes prior, then it was left unclassified by the GPS test and became the reference observation for the next test observation in the vehicle time-series.


The SVT was the final QC test which was applied to the observations that passed all previous tests. This test involved checking that the speed metadata for each observation was above a predetermined sensor ventilation threshold, $v_{sensor}$. Examining the speed-temperature pairs of the filtered dataset (not shown) revealed that the largest air temperatures (above $26^\circ$C) occurred for speeds below $25$km/h. We therefore set $v_{sensor} = 25$km/h. An observation passed the SVT if it had speed greater than $v_{sensor}$. Hence, any observations that were passed by the GPS test with low speeds were flagged by the SVT.

The QC-dataset contains $17425$ observations ($25.6\%$ of original dataset). A summary of the results of each QC test is provided in table \ref{ta: quality check passed and flagged numbers}. We note that the SIT and GPS test could not test every observation in the filtered dataset due to unavailable or unsuitable reference observations. The most discriminating test was the GPS test. The majority of observations flagged by the GPS test were likely the result of the $500$m update distance default setting on the app. We also note that the SVT was a fairly discriminating test.

The QC approach taken with this dataset relied upon range validity and time-series tests. For crowdsourced observations, time-series tests may be unsuitable as instrument identification metadata may be unavailable due to data privacy concerns. This may be overcome with appropriate encryption techniques \citep[e.g.,][]{v2x2019verheul} or by performing the QC locally on the sensing device \citep[e.g.,][]{hintz2019collectingprocessing}. Furthermore, the use of spatial consistency QC tests, which do not require instrument identification, would be a suitable replacement for time-series-based tests provided there is a sufficient density of observations in a given area \citep[e.g.,][]{nipen2019adopting}.

\begin{table}[]
    \centering
    \resizebox{\linewidth}{!}{\begin{tabular}{ccccc}
    \toprule
    QC test& Number of tested & Number of passed & Number of flagged & Number of untested \\
    & observations & observations & observations & observations \\
    \midrule
        
       Climatological range test & $32179$ & $32129$ & $50$ & $0$\\ 
       
       Stuck sensor test & $32179$ & $30124$ & $2008$ & $47$ \\ 
       
       GPS test & $32179$ & $20162$ & $11181$ & $836$ \\ 
       
       \midrule
       
       Sensor ventilation test & $19094$ & $17425$ & $1669$ & $0$ \\
      \bottomrule
    \end{tabular}}
    \caption{Summary of the results from all QC tests. The observations untested by the SIT and GPS test are due to a lack of reference observations. The observations passed by the SVT form the QC-dataset.}
    \label{ta: quality check passed and flagged numbers}
\end{table}

\section{Examination of the quality-controlled dataset} \label{sec: Examination of quality controlled dataset}

In this section we compare the QC-dataset with UKV model data, RWIS data, and MIDAS hourly data. Illustrative examples of the effect of sunny and rainy weather conditions on vehicle-based observations are presented in section \ref{sec: Effect of sunny and rainy weather conditions on vehicle-based observations}, analysis of observation-minus-background (OMB) and observation-minus-observation (OMO) statistics are discussed in section \ref{sec: Statistical analysis of observation-minus-background departures}, and vehicle-specific OMB statistics are examined in section \ref{sec: Vehicle specific observation-minus-background distributions}. We use UKV model data as the background in the OMB datasets and MIDAS hourly data in the OMO dataset.

The effects of different meteorological factors are quantified through statistical analysis of OMB departures grouped by sunny, cloudy and rainy weather conditions and season. The sunny dataset will consist of observations that occur between $09$:$00$ and $17$:$00$ on days with at least $6$ sunshine hours and less than $2$mm of rainfall. Therefore, the observations are likely to be influenced by solar radiation incident on UK roads. The rainy dataset will consist of observations that occur between $09$:$00$ and $17$:$00$ on days with at least $5$mm of rainfall and less than $2$ hours of sunshine. The cloudy dataset will consist of observations that occur between $09$:$00$ and $17$:$00$ on days with less $2$ hours sunshine and $2$mm rainfall. To obtain the weather-specific sub-datasets, we used the Met Office daily weather summaries \citep{metoffice2018dws}. The seasons we consider are Winter, defined as all data occurring between February 20th and March 20th 2018, and Spring, defined as all data occurring between March 21st and April 30th 2018. We note that these seasons do not conform to the usual definitions of meteorological winter and spring, but have been chosen due to the period of the Met Office trial and so that the Winter and Spring datasets each contain a similar number of observations.

\subsection{Case studies on the effect of sunny and rainy weather conditions on vehicle-based observations} \label{sec: Effect of sunny and rainy weather conditions on vehicle-based observations}

\begin{figure}
    \centering
    \includegraphics[width=\textwidth, trim = {5cm 0 5cm 0}]{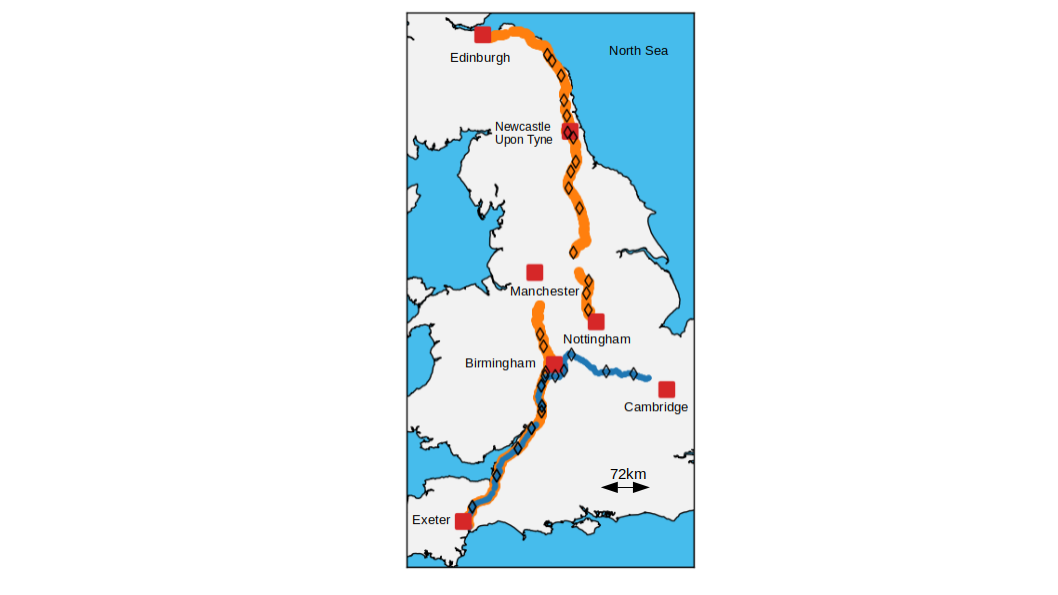}
    \caption{Map of the United Kingdom showing the location of the three time-series discussed in section \ref{sec: Effect of sunny and rainy weather conditions on vehicle-based observations}. The red squares show the location of cities passed by or near to the routes travelled in the three time-series. The black diamonds show the location of the RWIS stations passed on each journey. The two orange lines correspond to the sunny weather time-series and the blue line corresponds to the rainy weather time-series. The time-series S1 began near Exeter and travelled north towards Manchester. The time-series R1 travelled the same initial route as S1, but headed east from Birmingham towards Cambridge. The time-series S2 began in Edinburgh and travelled along the coast to Newcastle-upon-Tyne and then the vehicle travelled further inland and south towards Nottingham.}
    \label{fig: time-series map}
\end{figure}

\begin{figure}
    \centering
    \includegraphics[width=\textwidth]{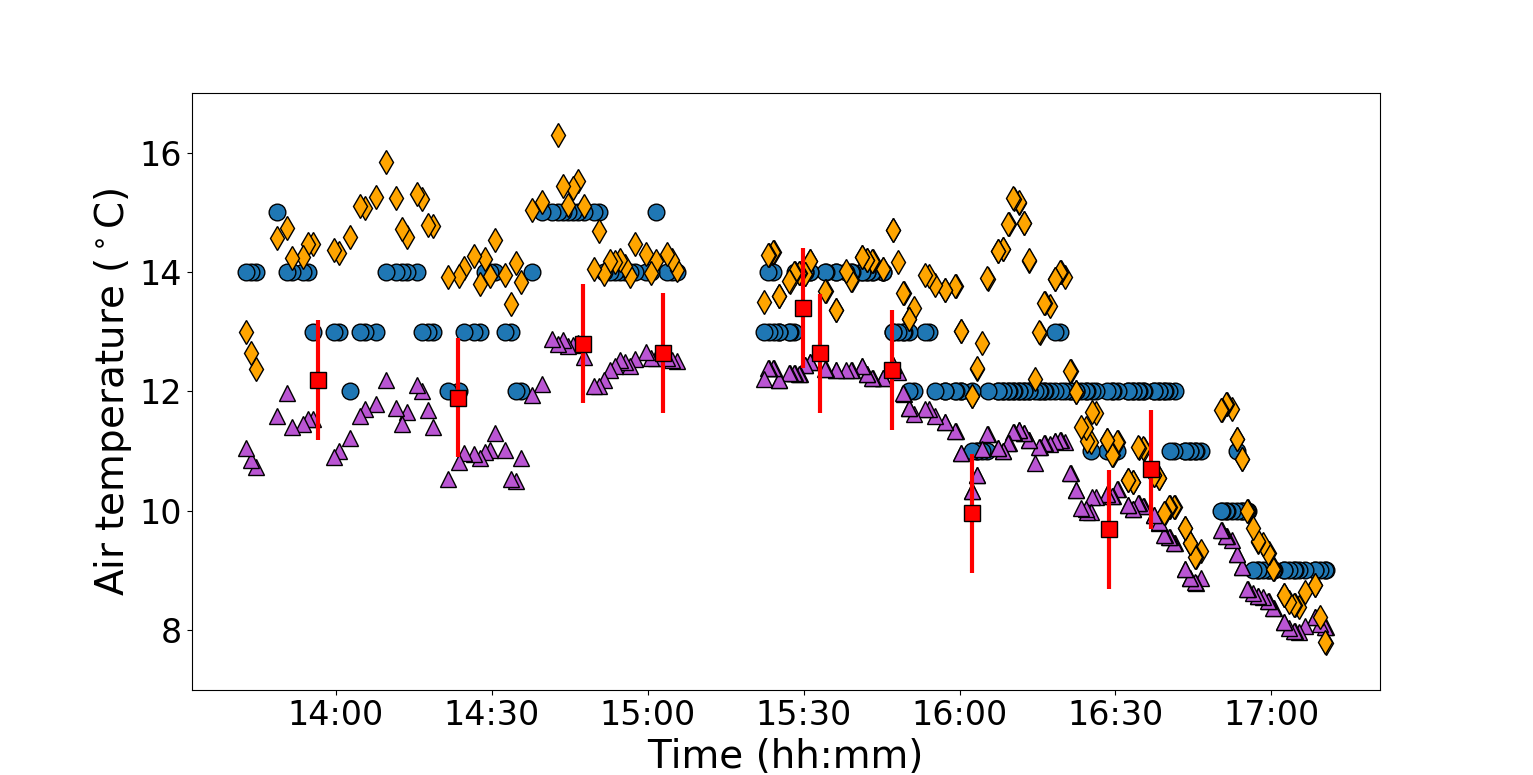}
    \caption{Time-series S1 of $212$ vehicle-based observations of air temperature (blue circles) from a single vehicle driving along the M5 motorway on 25th March 2018 during sunny weather. Also shown are UKV $1.5$m-air temperature (purple triangles) and UKV surface-air temperature (orange diamonds) linearly interpolated to the time and horizontal location of the vehicle observations, and RWIS observations of air temperature (red squares) linearly interpolated to the time the vehicle passed a station. The $1^\circ$C RWIS error bar represents the uncertainty used to assimilate RWIS observations into the UKV.}
    \label{fig: sunny time-series 0325}
\end{figure}

\begin{figure}
    \centering
    \includegraphics[width=\textwidth]{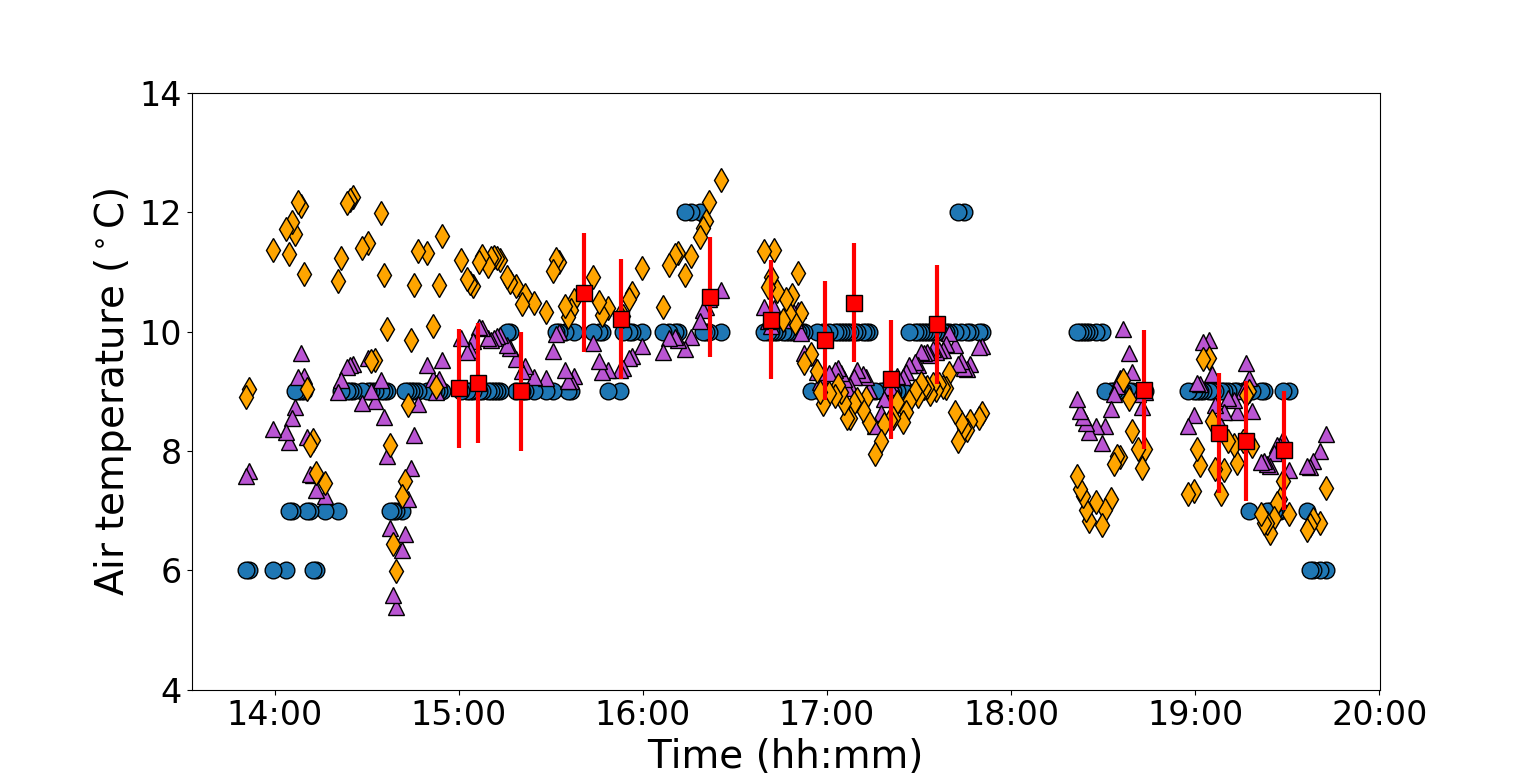}
    \caption{Time-series S2 of $193$ vehicle-based observations of air temperature (blue circles) from a single vehicle driving along the A1 and the M1 motorway on April 5th 2018 during sunny weather. Also shown are UKV $1.5$m-air temperature (purple triangles) and UKV surface-air temperature (orange diamonds) interpolated to the time and horizontal location of the vehicle observations, and RWIS observations of air temperature (red squares) interpolated to the time the vehicle passed a station. The $1^\circ$C RWIS error bar represents the uncertainty used to assimilate RWIS observations into the UKV.}
    \label{fig: sunny time-series 0405}
\end{figure}

\begin{figure}
    \centering
    \includegraphics[width=\textwidth]{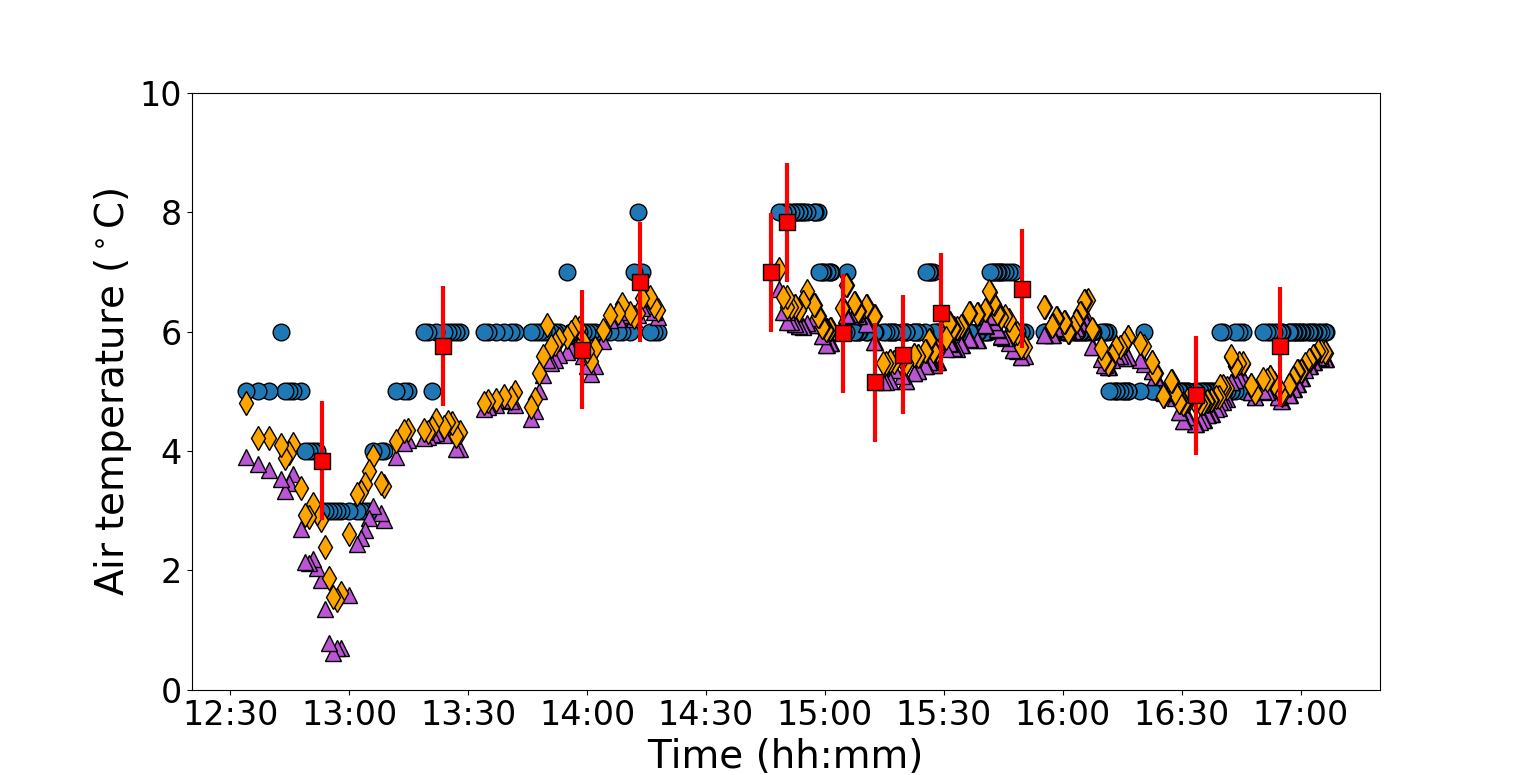}
    \caption{Time-series R1 of $259$ vehicle-based observations of air temperature (blue circles) from a single vehicle driving along the M5 and M42 motorways and the A5 on 30th March 2018 during rainy weather conditions. Also shown are UKV $1.5$m air temperature (purple triangles) and UKV surface air temperature (orange diamonds) interpolated to the time and horizontal location of the vehicle observations and RWIS observations of air temperature (red squares) interpolated to the time the vehicle passed a station. The $1^\circ$C RWIS error bar represents the uncertainty used to assimilate RWIS observations into the UKV.}
    \label{fig: rainy time-series 0330}
\end{figure}

We now show three time-series of vehicle-based observations of air temperature, $10$-minute UKV $1.5$m-air-temperature and surface-air-temperature model data, and RWIS observations of air temperature. The routes traversed in each time-series began and ended in suburban areas and were predominantly on major roads and major routes in rural areas which occasionally crossed urban areas. The location of each time-series is shown in figure \ref{fig: time-series map}. We denote the time-series shown in figure \ref{fig: sunny time-series 0325} as S1, figure \ref{fig: sunny time-series 0405} as S2, and figure \ref{fig: rainy time-series 0330} as R1. S1 and S2 are illustrative examples of the effect of sunny weather and R1 is an illustrative example of the effect of rainy weather on vehicle-based observations of air temperature. We note that the same vehicle produced the observations in S1 and R1, but a different vehicle produced the observations in S2. We also note that the large data gaps in the three time series are due to breaks in the journeys, and the two small data gaps in S2 are due to observations removed by the QC procedure.

For clarity, we will refer to the OMB data using $1.5$m-air temperature for the background as aOMB and using surface-air temperature data for the background as sOMB. Furthermore, we denote the bias (mean) and standard deviation of an aOMB dataset as $\mu_a$ and $\sigma_a$ respectively and the bias (mean) and standard deviation of a sOMB dataset as $\mu_s$ and $\sigma_s$ respectively.

\begin{table}
\begin{center}
\begin{tabular}{cccc}
    \hline
    & \multicolumn{3}{c}{Time-series} \\
    Summary Statistics & S1 & S2 & R1 \\
    \hline
    Number of observations & $212$ & $193$ & $259$ \\
    $\mu_a$ $^{\circ}$C & $1.44$ & $0.02$ & $0.65$ \\
    $\sigma_a$ $^{\circ}$C & $0.71$ & $0.90$ & $0.71$ \\
    $\mu_s$ $^{\circ}$C & $-0.46$ & $-0.27$ & $0.37$ \\
    $\sigma_s$ $^{\circ}$C & $1.21$ & $1.71$ & $0.64$ \\
    \hline
\end{tabular}
\end{center}
\caption{Summary of the OMB statistics for the three time-series shown in figure \ref{fig: time-series map} using UKV $1.5$m-air temperature and surface-air temperature as the background. The uncertainty in the mean for each time-series is less than $0.1^{\circ}$C.}
\label{ta: time-series statistics summary}
\end{table}

Figure \ref{fig: sunny time-series 0325} shows data from sunny weather conditions on March 25th 2018, including the time-series S1, UKV and RWIS data. The OMB summary statistics for S1 are shown in table \ref{ta: time-series statistics summary}. The sun rose at $06$:$52$ and set at $18$:$22$ on this day. The RWIS stations included in this time-series recorded a dry road-state when the vehicle passed the station. As a result of these conditions, we expect the sensible heat flux emitted by the road to be large and the road surface to have a noticeable heating effect on the air temperature above (see section \ref{sec: Representation error}). Additionally, we expect the surface-air temperature to be larger than the $1.5$m-air temperature as the sensible heat flux emitted by the UKV surface will also be large. The mean difference between the interpolated RWIS observations and the nearest-in-time UKV model data reveals that RWIS observations are in most agreement with $1.5$m-air temperature and in least agreement with surface-air temperature. There is a clear separation between UKV $1.5$m-air temperature and surface-air temperature at the start of the time-series that gradually decreases as the net radiation absorbed by the UKV surface decreases. The vehicle-based observations generally lie between the model fields as seen by the difference in sign between the biases, $\mu_a$ and $\mu_s$. The vehicle-based observations on average agree most with surface-air temperature as $|\mu_s| < |\mu_a|$. This is consistent with the height of the vehicle sensor which is likely to be between the model field heights of $0$m and $1.5$m but closer to $0$m than $1.5$m. Calculating the standard deviation of the sOMB and aOMB departures shows that the sOMB departures are more variable as $\sigma_a < \sigma_s$. We hypothesise that the variability of the UKV sensible heat flux induced by the sunny weather conditions is the mechanism responsible for the larger sOMB variability. 

Figure \ref{fig: sunny time-series 0405} shows data from sunny weather conditions on April 5th 2018, including the time-series S2, UKV and RWIS data. The OMB summary statistics for S2 are shown in table \ref{ta: time-series statistics summary}. The sun rose at $05$:$27$ and set at $18$:$40$ on this day. The RWISs included in this time-series recorded a dry road-state when the vehicle passed the station. Similarly to the data in figure \ref{fig: sunny time-series 0325}, the RWIS observations are in most agreement with the UKV $1.5$m-air temperature and in least agreement with surface-air temperature. The surface-air temperature is larger than the $1.5$m-air temperature for the first half of this time-series. From approximately $17$:$00$ we see that $1.5$m-air temperature is greater than surface-air temperature. We hypothesise that this is due to the stabilisation of the boundary layer \citep{stull1988introduction2}. In contrast to S1, the vehicle-based observations are closest to UKV $1.5$m-air temperature at the beginning of the time-series even though the sensible heat flux emitted from the road surface is expected to be greatest during this period. Possible reasons for this include cool breezes from the North Sea influencing the vehicle-based observations during the beginning of the time-series (see route map in figure \ref{fig: time-series map}) or because the air temperature is measured by a different vehicle's instrument. Furthermore, the difference between the biases $\mu_a$ and $\mu_s$ is large for S1 and small for S2. This, however, is likely due to the large number of observations that occurred during the evening for S2 when the temperature gradient between the surface and $1.5$m is expected to be small. Considering the observations from the first 3 hours of S2 only, when the net solar radiation absorbed by the road and UKV surface is expected to be large, we find that the difference between the biases $\mu_a$ and $\mu_s$ is more profound. We note that the standard deviations $\sigma_a$ and $\sigma_s$ are larger for S2 than S1. This is likely due to the following two reasons. The first reason is the relatively long temporal length of the S2 time-series. The second reason is the possible transition to the nocturnal boundary layer as the dynamics induced by solar heating and the generation of convective plumes begins to cease and surface layer starts to become stably stratified. However, it is also plausible that the placement of the external air-temperature sensors on the two vehicles is contributing to this behaviour.


Figure \ref{fig: rainy time-series 0330} shows data from rainy weather conditions on March 30th 2018, including the time-series R1, UKV and RWIS data. The OMB summary statistics for R1 are shown in table \ref{ta: time-series statistics summary}. The RWIS stations included in this time-series recorded either a wet road-state or trace amounts of water at the road surface when the vehicle passed the station. As a result of these conditions, we expect the sensible heat emitted by the road to be small and the road surface to have a reduced effect on the air temperature above (see section \ref{sec: Representation error}). We note that the drop in air temperature between $12$:$30$ and $13$:$30$ is caused by an increase in altitude and an occluded front. The mean difference between the interpolated RWIS observations and the nearest-in-time UKV data reveals that RWIS observations are now in greater agreement with surface-air temperature than $1.5$m-air temperature. The two UKV model fields are similar throughout the time-series with multiple segments where the vehicle-based observations are greater than both fields. The vehicle-based observations are on average greater than the UKV model data as the biases $\mu_a, \mu_s > 0^{\circ}$C but agree more with surface-air temperature as $\mu_a > \mu_s$. This indicates that there are additional factors affecting the vehicle-based observations. Potential explanations for this behaviour are given in section \ref{sec: Discussion of bias and variability exhibited by the OMB datasets}. We note that while the aOMB departures are more variable than the sOMB departures (i.e. $\sigma_a > \sigma_s$), they are similar in size. 

The effect of the sensible heat emitted by the road and UKV surfaces can be observed through comparison of the S1 and R1 time-series shown in figures \ref{fig: sunny time-series 0325} and \ref{fig: rainy time-series 0330}, respectively. In sunny weather, the sensible heat emitted by the road and UKV surface will be large, resulting in a stronger vertical air-temperature gradient between the surface and the $1.5$m height. In rainy weather, the sensible heat emitted by the road and UKV surface will be small, leading to a vertical air-temperature profile that is more uniform. Hence, the difference between the biases $\mu_a$ and $\mu_s$ will be larger in sunny weather conditions than rainy weather conditions. The OMB standard deviations calculated for each time-series show a negligible difference for $\sigma_a$ and a noticeable difference for $\sigma_s$ between the two time-series. For the sOMB standard deviation $\sigma_s$, we see that it is smaller for rainy weather and larger for sunny weather. This is likely because the variability of the sensible heat emitted by the UKV surface will be greater in sunny weather than rainy weather. However, there may be other contributing factors such as the difference in observation operator error between the two time-series.

\subsection{Statistical analysis of observation-minus-background and observation-minus-observation departures} \label{sec: Statistical analysis of observation-minus-background departures}

In this section we investigate the uncertainty present in the QC-dataset through statistical analysis of OMB and OMO departures. The OMB datasets will be partitioned into weather-specific and seasonal sub-datasets so that we may examine how the OMB uncertainty changes with weather conditions and season. As there are only $347$ observations within $1.5$km of a MIDAS station, we will not split the OMO dataset into weather-specific and seasonal sub-datasets. We now discuss the characteristics of each dataset. 


\subsubsection{QC-dataset OMB and OMO statistics} \label{sec: QC-dataset}

The OMB and OMO statistics corresponding to the QC-dataset are given in table \ref{ta: OMB statistics summary}. Examining the OMB statistics shows that the vehicle-based observations are in poorer agreement with $1.5$m-air temperature than surface-air temperature as the biases satisfy $| \mu_a | > | \mu_s |$. This is expected as external air-temperature sensors likely measure air temperature nearer to the surface than to a height of $1.5$m. As $\mu_a > 0^\circ$C and $\mu_s > 0^\circ$C, vehicle-based observations are on average warmer than both UKV model fields despite measuring the air temperature between them. Possible reasons for this behaviour are discussed in section \ref{sec: Discussion of bias and variability exhibited by the OMB datasets}. For the standard deviations, we have that $\sigma_a < \sigma_s$, showing that the sOMB dataset is more variable than the aOMB dataset. This is also visible in the aOMB and sOMB distributions shown by the histograms in figure \ref{fig: OMB for qc and weather specific observations}a. While the distributions overlap substantially, a higher peak is seen for the aOMB distribution, whereas the sOMB distribution has a larger left tail. It is likely that the background uncertainty of surface-air temperature is greater than $1.5$m-air temperature due to the simplifying assumptions made by the UKV in modelling the SEB of a grid box (see section \ref{sec: NWP model data}). 

Examining the OMO statistics reveals that vehicle-based observations are on average warmer than MIDAS observations. Comparing the OMO departure bias, $\mu_m$, with the biases $\mu_a$ and $\mu_s$ obtained from the QC-dataset OMB departures, we find that $\mu_s < \mu_m < \mu_a$. Hence, the vehicle-based observations on average agree more with MIDAS data than $1.5$m-air temperature but still agree most with surface-air temperature. While it is plausible that vehicle-based observations will generally agree more with MIDAS data than the UKV $1.5$m-air temperature model data, we note the following two issues with these calculations. Firstly, the MIDAS data were not interpolated to the location of the vehicle-based observations and there will likely be differences in elevation between the two. As air temperature is expected to change with elevation in the surface layer \citep{stull1988introduction}, the omission of any vertical interpolation may cause a bias. Secondly, while the variability of the OMO dataset is less than the variability of any OMB dataset shown in table \ref{ta: OMB statistics summary}, it is calculated with far fewer observations making the statistic less reliable.

\subsubsection{Weather-specific OMB datasets} \label{sec: Weather-specific datasets}

The OMB statistics for each weather-specific dataset are given in table \ref{ta: OMB statistics summary}. For all weather types, the bias $\mu_a$ is positive showing that vehicle-based observations are on average warmer than UKV $1.5$m-air temperature regardless of weather conditions. For the sunny and cloudy datasets, the bias $\mu_s$ is negative, whereas for the rainy dataset $\mu_s$ is positive. This agrees with the results of the three time-series discussed in section \ref{sec: Effect of sunny and rainy weather conditions on vehicle-based observations}. This also suggests that the vehicle-based observations studied in this paper may be colder on average than UKV surface-air temperature in dry conditions. The smallest differences between the biases $\mu_a$ and $\mu_s$ occurs for the rainy dataset while the largest difference occurs for the sunny dataset which is also seen in the S1 and R1 time-series discussed in section \ref{sec: Effect of sunny and rainy weather conditions on vehicle-based observations}. For the rainy dataset biases, we have that $\mu_a > 0^\circ$C and $\mu_s > 0^\circ$C which indicates the vehicle-based observations are on average warmer than the two UKV model fields. This suggests that there are other influencing factors on the vehicle-based observations that are not represented in the UKV as vehicles measure the air temperature between these two heights. Potential explanations for this behaviour are discussed in section \ref{sec: Discussion of bias and variability exhibited by the OMB datasets}.

\begin{figure}
  \begin{subfigure}[t]{.49\textwidth}
    \centering
    \includegraphics[width=\linewidth]{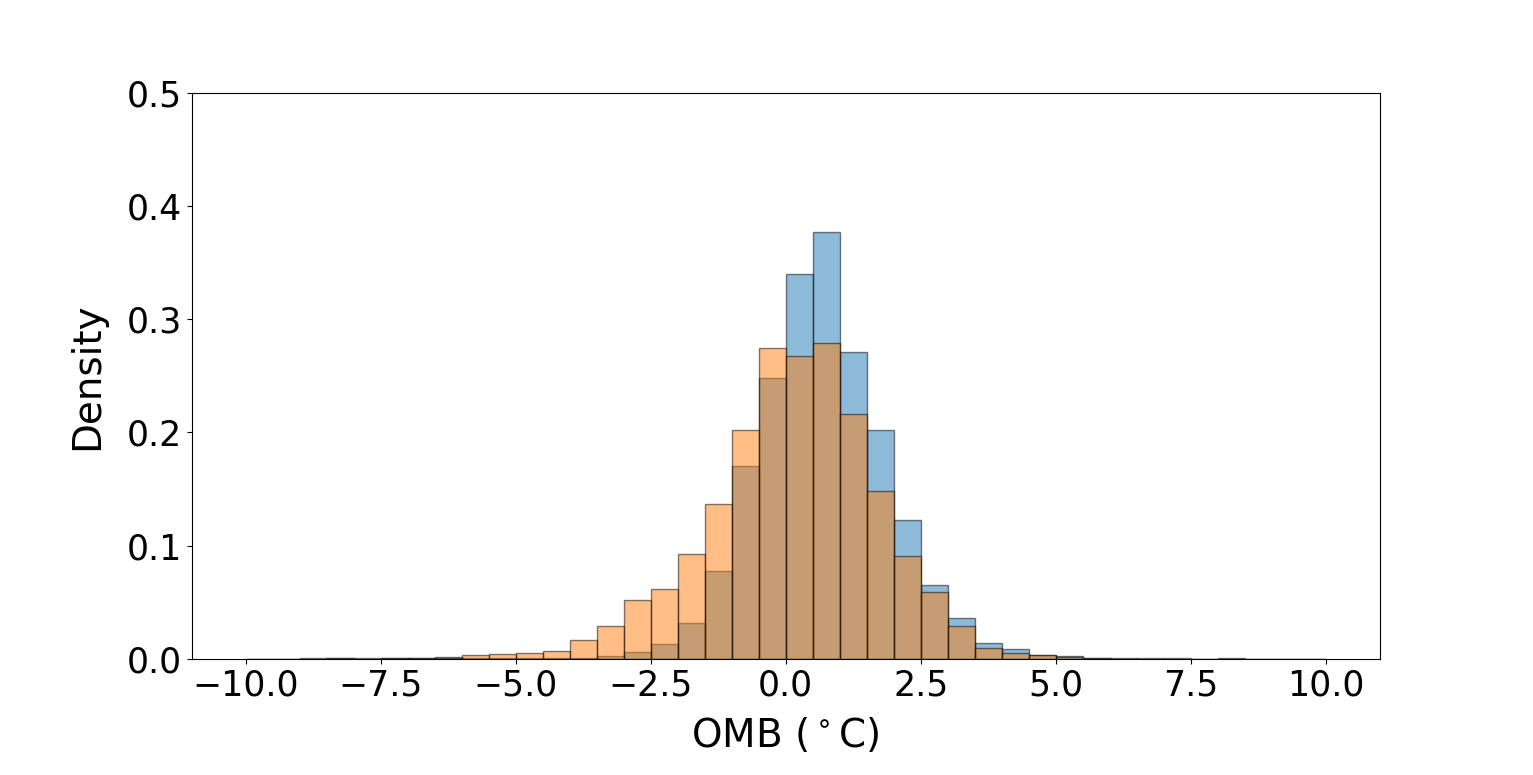}
    \caption{QC-dataset}
  \end{subfigure}
  \hfill
  \begin{subfigure}[t]{.49\textwidth}
    \centering
    \includegraphics[width=\linewidth]{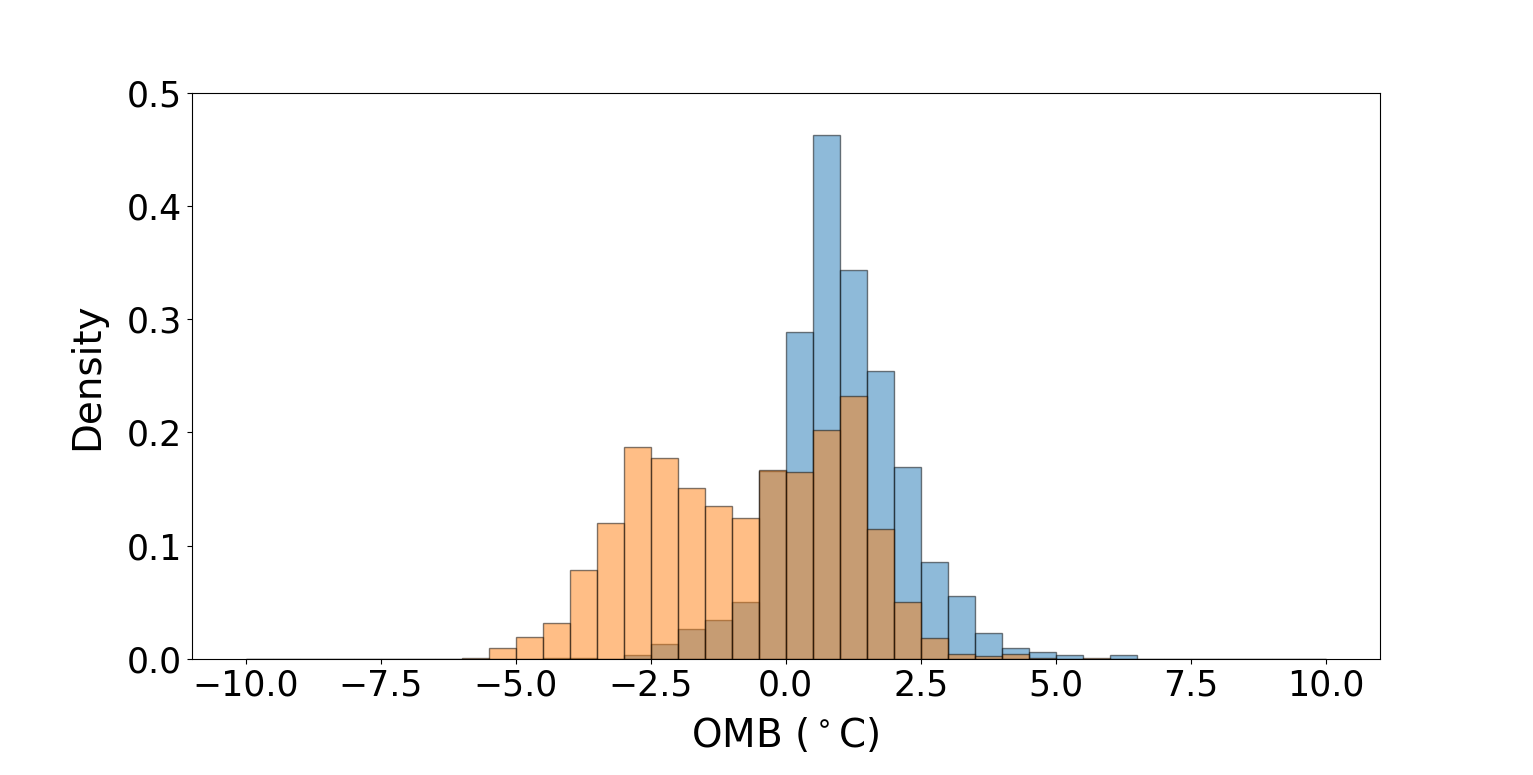}
    \caption{Sunny dataset}
  \end{subfigure}

  \medskip

  \begin{subfigure}[t]{.49\textwidth}
    \centering
    \includegraphics[width=\linewidth]{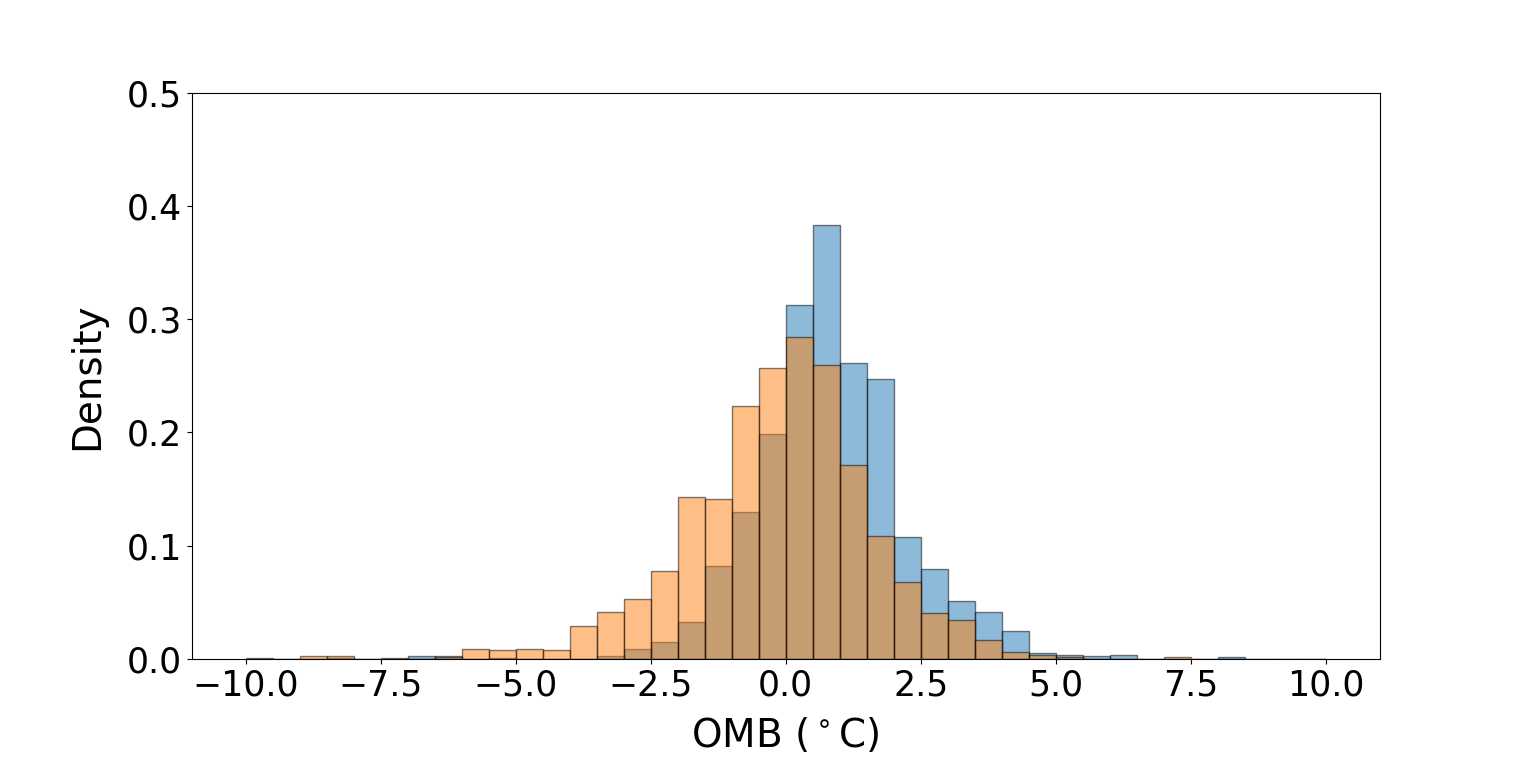}
    \caption{Cloudy dataset}
  \end{subfigure}
  \hfill
  \begin{subfigure}[t]{.49\textwidth}
    \centering
    \includegraphics[width=\linewidth]{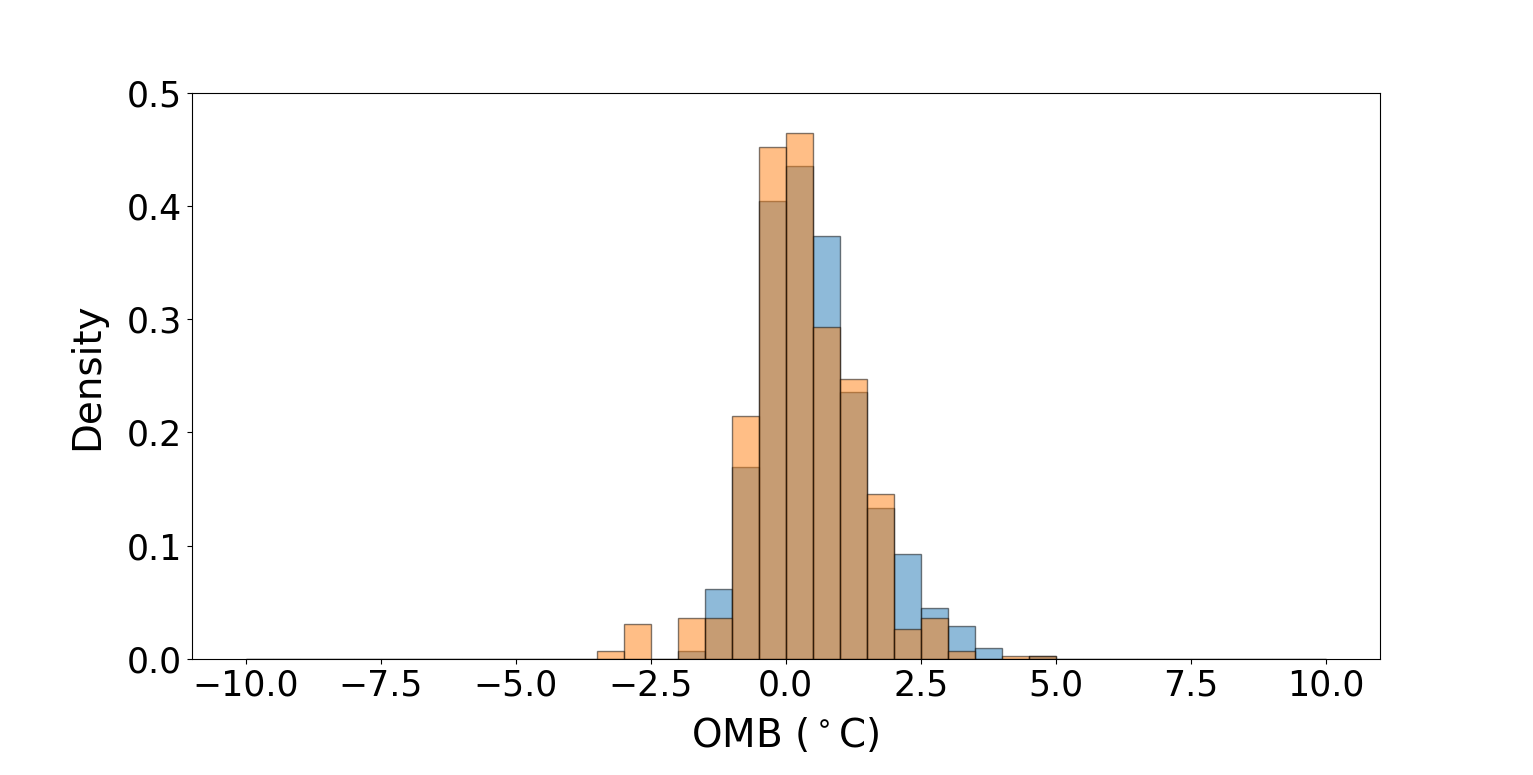}
    \caption{Rainy dataset}
  \end{subfigure}
 
 \medskip
 
 \begin{subfigure}[t]{.49\textwidth}
    \centering
    \includegraphics[width=\linewidth]{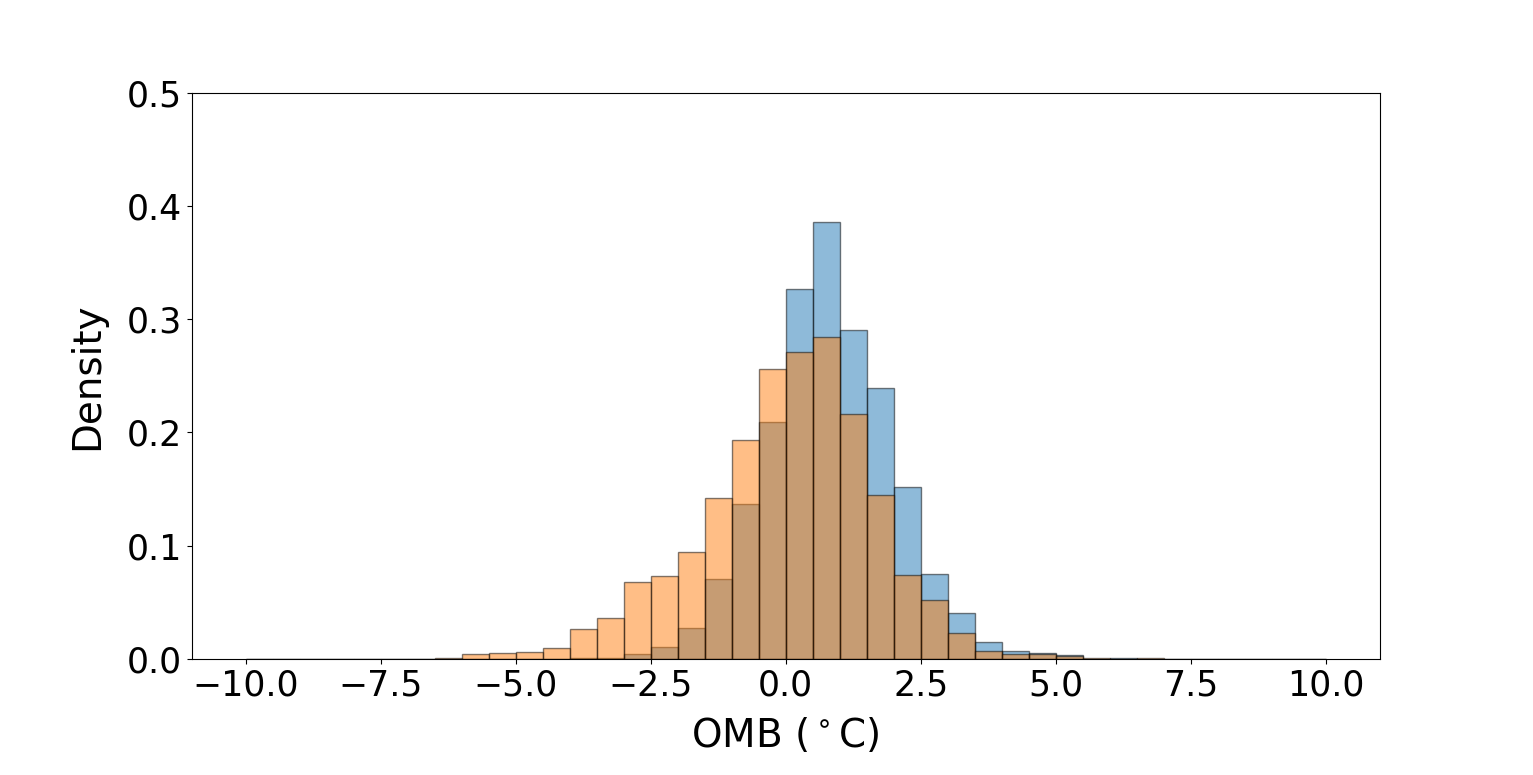}
    \caption{Spring dataset}
  \end{subfigure}
  \hfill
  \begin{subfigure}[t]{.49\textwidth}
    \centering
    \includegraphics[width=\linewidth]{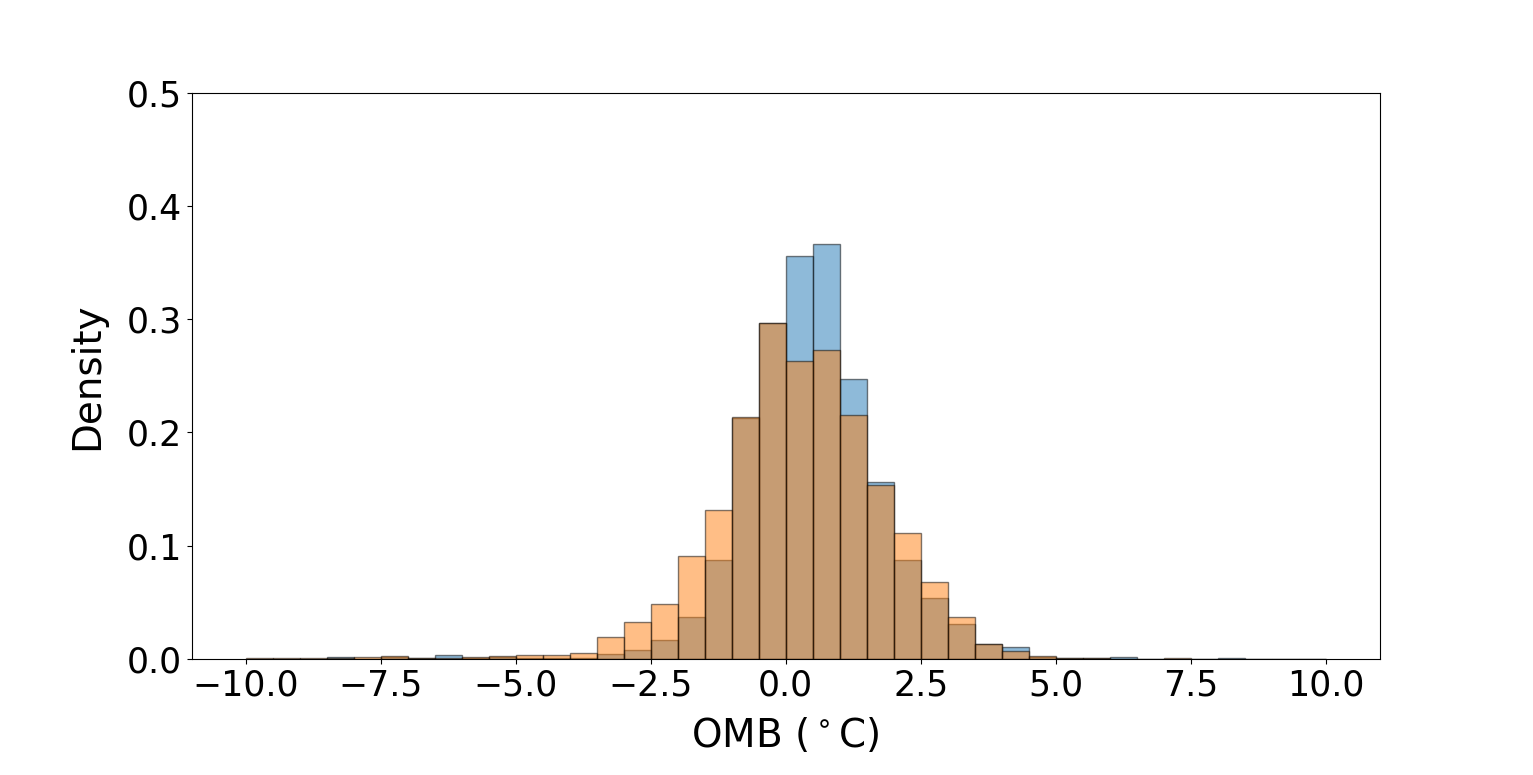}
    \caption{Winter dataset}
  \end{subfigure}
 
 \caption{OMB histograms datasets corresponding to the datasets in table \ref{ta: OMB statistics summary}. Bins of width $0.5^\circ$C have been used for each histogram. The blue bars correspond to the aOMB bins and the orange bars correspond to the sOMB bins.}
 
 \label{fig: OMB for qc and weather specific observations}
\end{figure}

Inspection of the weather-specific standard deviations reveals that, for sunny and cloudy weather conditions, $\sigma_s$ is noticeably larger than $\sigma_a$. This difference in variability is shown in the histograms for the sunny dataset (figure \ref{fig: OMB for qc and weather specific observations}b) and the cloudy dataset (figure \ref{fig: OMB for qc and weather specific observations}c). The sunny aOMB distribution is unimodal and the sunny sOMB distribution is bimodal. The bimodal structure may be due to intermittent cloud cover on the days with less sunshine hours or the relatively small size of the sunny dataset. Similarly to the QC-dataset, the cloudy aOMB and sOMB distributions overlap substantially, but a higher peak is seen for the aOMB distribution and a larger left tail is seen for the sOMB distribution. For rainy weather conditions, the standard deviations $\sigma_a$ and $\sigma_s$ are similar leading to similar aOMB and sOMB distributions as shown in figure \ref{fig: OMB for qc and weather specific observations}d. 

Overall, the variability for the weather-specific datasets agrees with the variability calculated for the time-series discussed in section \ref{sec: Effect of sunny and rainy weather conditions on vehicle-based observations}. We also find that the standard deviation $\sigma_s$ is larger for the sunny sOMB dataset than for the cloudy sOMB dataset. This behaviour is likely the result of the increased variability of the sensible heat emitted by roads and the UKV surface during sunny weather conditions due to the larger amount of solar radiation absorbed by the two surfaces. This also suggests that the uncertainty of vehicle-based observations may be greatest in sunny weather conditions due to the combination of the radiative effects on the vehicle sensor and the representation uncertainty. Conversely, the aOMB standard deviation $\sigma_a$ is largest for the cloudy dataset and not the sunny dataset. This, however, may be due to changes in the sky-view factor due to variable cloud cover or because the cloudy dataset has more observations than the sunny dataset. We hypothesise that the rainy dataset standard deviations $\sigma_s$ and $\sigma_a$ are similar for the following three reasons. Rain increases the availability of water at the UKV surface which reduces the emitted sensible heat flux and hence the vertical air-temperature profile will be more uniform. There is also little to no sun at the times and locations of the vehicle-based observations in the rainy dataset resulting in negligible radiation reflected by the road surface incident on the vehicle temperature sensor. Finally, the rainy dataset is the smallest of our three weather-specific datasets and so the OMB statistics are the least robust.

\subsubsection{Seasonal OMB datasets} \label{sec: Seasonal datasets}

The OMB statistics for the seasonal datasets are given in table \ref{ta: OMB statistics summary} and the histograms are plotted in figures \ref{fig: OMB for qc and weather specific observations}e (Spring) and \ref{fig: OMB for qc and weather specific observations}f (Winter). We include information on the seasonal datasets to provide a baseline for the vehicle-specific analysis in section \ref{sec: Vehicle specific observation-minus-background distributions}. For the seasonal datasets, the vehicle-based observations are on average greater than the UKV model fields except for surface-air temperature in Spring where they are approximately the same. Comparing the biases of the seasonal datasets we see that $\mu_s$ is smaller and $\mu_a$ is greater in Spring than in Winter. Inspection of the standard deviations of the seasonal datasets reveals that $\sigma_s$ is larger than $\sigma_a$ in both Winter and Spring. Comparing the seasonal OMB statistics, we find that the sOMB standard deviation $\sigma_s$ is larger for the Spring dataset than for the Winter dataset. This agrees with the results of the weather-specific datasets as the Spring dataset contains more sunny days than the Winter dataset.

\subsubsection{Discussion of the uncertainty exhibited by the OMB datasets} \label{sec: Discussion of bias and variability exhibited by the OMB datasets}

In this section we discuss several possible contributions to the uncertainty exhibited in the OMB statistics shown in table \ref{ta: OMB statistics summary}.
\begin{itemize}
    \item 
    The vehicle-based observations of air temperature are precise to $1^{\circ}$C. The details of the observation processing by the OBD system and app are unknown. However, an indication of the expected size of the processing errors can be obtained by considering the quantization error from a typical procedure that rounds to the nearest integer. In this case, the root-mean-squared quantization error is $\sqrt{1/12}$ $^\circ$C and may be a positive or negative error  \citep{widrow1996statistical}.
    \item
    As discussed in section \ref{sec: Instrument errors}, the external air-temperature sensors may exhibit a warm bias from extraneous sources. For instance, the sensor may be in close proximity to the vehicle engine or the location of the sensor may be inadequate for sensor ventilation. A rough estimate of the bias due to the vicinity of the engine is $5^\circ$C - $25^\circ$C \citep{mercelis2021email}. However there are several other factors that may be influencing this estimate (e.g. differing vehicles).
    \item
    In unstable atmospheric conditions, the vertical air-temperature gradient between the surface and the $1.5$m height will be negative. When the sensible heat emitted by the road and UKV surfaces is large, the vertical air-temperature gradient will be large. Therefore, surface-air temperature will be warmer on average than vehicle-based observations which likely measure air temperature between $20$cm and $100$cm above the road surface. Similarly, $1.5$m-air temperature will be cooler on average than vehicle-based observations. 
    \item
    The vehicle-based observations have not been corrected to the elevation of the model grid box. In NWP, it is common to correct surface observations using a standard adiabatic lapse rate of $0.0065^\circ$C m$^{-1}$ \citep{dutra2020environmental,cosgrove2003real}.
    \item
    The road surface temperature is highly variable due to sky-view factors and traffic \citep{chapman2011spatial}. As noted in section \ref{sec: Representation error}, these factors can change the local temperature by as much as $2$ or $3^\circ$C.
    \item
    The difference in sensible heat emitted by the road surface and the UKV surface may contribute to the OMB biases. As discussed in section \ref{sec: NWP model data}, these model errors will vary depending on the land-surface (e.g. urban/rural) but are difficult to quantify. 
\end{itemize}


\begin{table}
\begin{center}
\begin{tabular}{ccccc}
    \hline
    \multicolumn{5}{c}{OMB and OMO statistics} \\
     Dataset & Number of observations & Departure & Mean ($^{\circ}$C) & Standard deviation ($^{\circ}$C) \\
    \hline
    \multirow{3}{*}{QC-dataset} & \multirow{2}{*}{$17425$} & aOMB & $0.67$ & $1.24$\\
    & & sOMB & $0.12$ & $1.59$\\
    & $347$ & OMO & $0.29$ & $0.95$ \\
    \hline
    \multirow{2}{*}{Sunny} & \multirow{2}{*}{$1878$} & aOMB & $1.05$ & $1.15$\\
    & & sOMB & $-0.77$ & $1.86$\\
    \hline
    \multirow{2}{*}{Cloudy} & \multirow{2}{*}{$2366$} & aOMB & $0.84$ & $1.41$\\
    & & sOMB & $-0.14$ & $1.74$\\
    \hline
    \multirow{2}{*}{Rainy} & \multirow{2}{*}{$840$} & aOMB & $0.56$ & $0.97$\\
    & & sOMB & $0.30$ & $1.00$\\
    \hline
    \multirow{2}{*}{Winter} & \multirow{2}{*}{$7798$} & aOMB & $0.48$ & $1.29$\\
    & & sOMB & $0.26$ & $1.55$\\
    \hline
    \multirow{2}{*}{Spring} & \multirow{2}{*}{$9627$} & aOMB & $0.82$ & $1.17$\\
    & & sOMB & $0.01$ & $1.61$\\
    \hline
\end{tabular}
\end{center}
\caption{Summary of the OMB and OMO  departure statistics for each dataset. The uncertainty in the mean for each dataset is less than $0.1^\circ$C.}
\label{ta: OMB statistics summary}
\end{table}

\subsection{Vehicle specific observation-minus-background departure distributions} \label{sec: Vehicle specific observation-minus-background distributions}

\begin{sidewaysfigure}
    \centering
    \includegraphics[width=\textwidth]{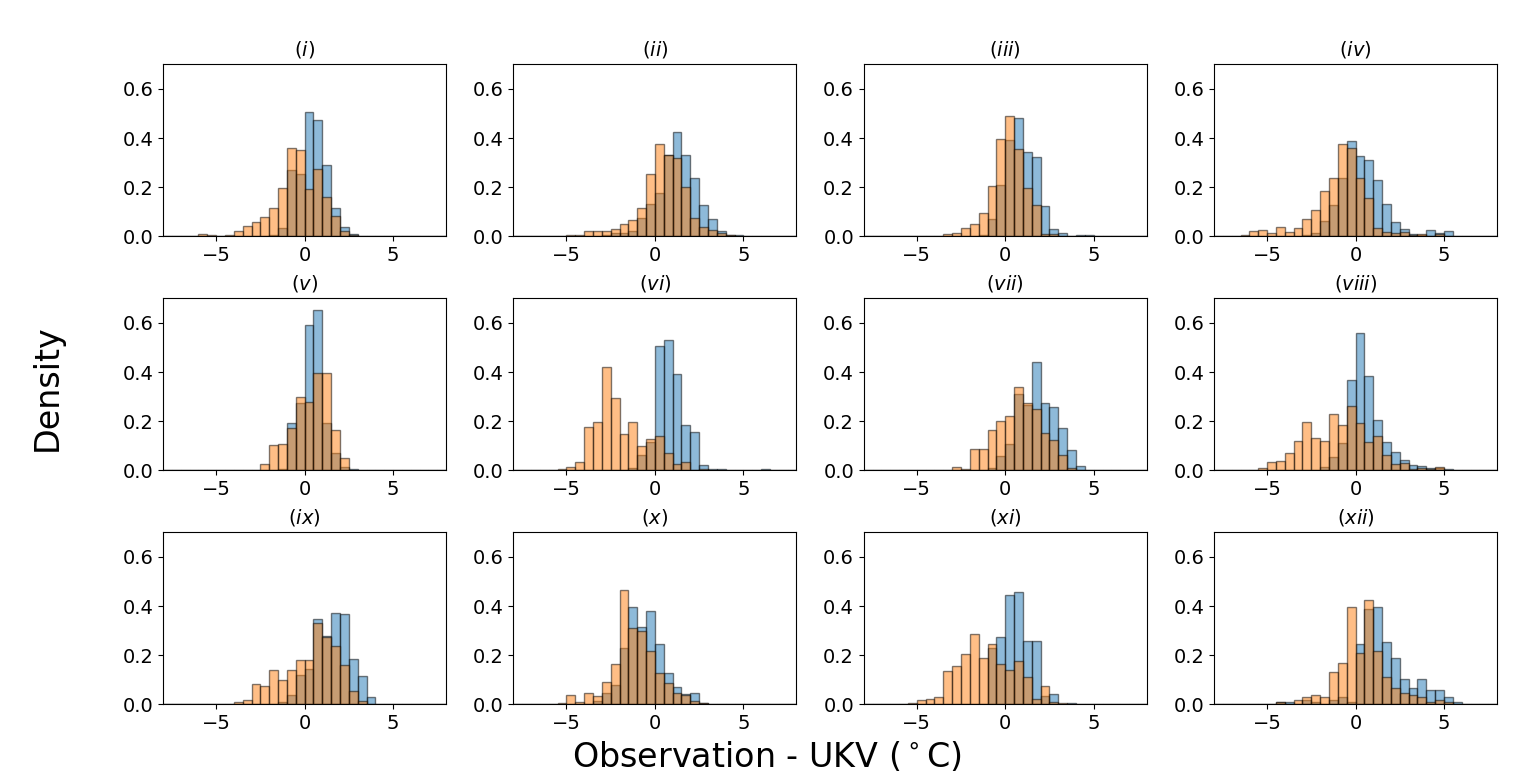}
    \caption{OMB histograms for the $12$ vehicles with the most observations between $9$am and $5$pm in the QC-dataset. The aOMB distributions are shown by the blue bins, the sOMB distributions are shown by the orange bins, and the overlap in distributions is shown by the composite. The same bins of width $0.5^\circ$C have been used for each histogram. Information on the statistics and observation meteorological conditions for each vehicle is given in table \ref{ta: vehicle OMB statistics summary}.}
    \label{fig: vehicle specific omb histograms}
\end{sidewaysfigure}

Throughout the Met Office trial, $31$ vehicles were used to produce vehicle-based observations. To investigate whether error statistics differ with vehicle, we plot OMB histograms for each of the $12$ vehicles with the most observations between $09$:$00$ and $17$:$00$ in the QC-dataset in figure \ref{fig: vehicle specific omb histograms}. We use only observations between $09$:$00$ and $17$:$00$ so that the boundary layer is likely to be unstable (i.e. UKV surface-air temperature is greater than the $1.5$m-air temperature). This is to avoid the complications of interpreting the OMB statistics experienced with the S2 time-series discussed in section \ref{sec: Effect of sunny and rainy weather conditions on vehicle-based observations} and so that more observations can be classified into the weather-specific data types discussed in section \ref{sec: Statistical analysis of observation-minus-background departures}. The OMB statistics are summarised in table \ref{ta: vehicle OMB statistics summary}. Also included in table \ref{ta: vehicle OMB statistics summary} is the percentage of observations for each vehicle occurring in each weather-specific and seasonal dataset discussed in section \ref{sec: Statistical analysis of observation-minus-background departures}. We note that it is difficult to draw definitive conclusions in this examination for two reasons. Firstly, many of the vehicles experience different weather conditions and there are many observations which we are unable to classify into a weather type. Secondly, there are only $12$ vehicles with an acceptable number of observations that we can examine.

The majority of histograms shown in figure \ref{fig: vehicle specific omb histograms} resemble normal distributions. The aOMB and sOMB distributions of vehicles (ii), (iii), (iv), (x), and (xii) are qualitatively similar with the visual distinction between them a result of the difference in means. The remaining vehicles have noticeably different aOMB and sOMB distributions.

Examining the biases of the vehicle-specific OMB distributions shows that there are some vehicles which agree more with UKV $1.5$m air temperature than surface air temperature as $| \mu_a | < | \mu_s |$. For these vehicles it is possible that the external air temperature sensor is located closer to $1.5$m height than the road surface or there are additional unknown factors affecting the vehicle-based observations as in S2 discussed in section \ref{sec: Effect of sunny and rainy weather conditions on vehicle-based observations}. The values of the biases $\mu_a$ and $\mu_s$ also vary substantially between vehicles. The sign of $\mu_s$ also varies with vehicle whereas only vehicle (x) has negative $\mu_a$ which suggests that some element of the vehicle's external air-temperature sensor or processing procedure may have a cold bias. 

Except for vehicle (xii), we find that the aOMB dataset is less variable than the sOMB dataset (i.e. $\sigma_a < \sigma_s$) for all vehicles which agrees with the results obtained in sections \ref{sec: Effect of sunny and rainy weather conditions on vehicle-based observations} and \ref{sec: Statistical analysis of observation-minus-background departures}. The values of the standard deviations $\sigma_a$ and $\sigma_s$ also vary substantially between vehicles. When stratifying the vehicles by seasonal contribution, we are able to find some agreement between the OMB dataset variability. This can be seen between the groups of vehicles (ii) and (vii), vehicles (iv), (viii) and (x), and vehicles (i) and (xi). Each vehicle in these groups contains similar ratios of Winter to Spring observations. For vehicles (iii), (vi), (ix) and (xii), which contain predominantly Spring observations, we see that the aOMB standard deviation $\sigma_a$ is similar except for vehicle (xii) and the sOMB standard deviation $\sigma_s$ is similar except for vehicle (iii). This shows that vehicles may have similar OMB uncertainty if they have similar ratios of seasonal observations.

\setlength{\tabcolsep}{12pt}
\begin{center}
\begin{sidewaystable}
\begin{tabularx}{1.0\linewidth}{ c*{12}{C} }
    \toprule
    & \multicolumn{12}{c}{Vehicle} \\
    Summary statistics & (i) & (ii) & (iii) & (iv) & (v) & (vi) & (vii) & (viii) & (ix) & (x) & (xi) & (xii)\\
\midrule
Number of observations   &  $1533$   &  $1522$  &  $910$  &   $904$ &   $780$ &  $593$ &  $562$  &  $502$ &  $497$  &   $421$ &  $342$ &   $212$ \\
$\mu_a$ $^\circ$C & $0.43$ & $1.24$ & $0.89$ & $0.40$ & $0.42$ & $0.91$ & $1.80$ & $0.59$ & $1.51$ & $-0.56$ & $0.61$ & $1.56$ \\
$\sigma_a$ $^\circ$C & $0.90$ & $1.06$ & $0.92$ & $1.14$ & $0.79$ & $0.91$ & $1.00$ & $1.03$ & $1.00$ & $1.04$ & $0.91$ & $1.23$ \\
$\mu_s$ $^\circ$C & $-0.42$ & $0.44$ & $0.14$ & $-0.84$ & $0.36$ & $-1.87$ & $0.78$ & $-0.94$ & $0.29$ & $-1.17$ & $-1.03$ & $0.40$\\
$\sigma_s$ $^\circ$C & $1.17$ & $1.19$ & $0.98$ & $1.25$ & $1.02$ & $1.21$ & $1.15$ & $1.36$ & $1.22$ & $1.14$ & $1.26$ & $1.20$ \\
\midrule
Sunny percentage & $0\%$ & $23\%$ & $22\%$ & $3\%$ & $34\%$ & $64\%$ & $13\%$ & $9\%$ & $36\%$ & $16\%$ & $22\%$ & $43\%$ \\
Cloudy percentage & $25\%$ & $23\%$ & $44\%$ & $18\%$ & $7\%$ & $10\%$ & $31\%$ & $26\%$ & $12\%$ & $47\%$ & $35\%$ & $38\%$ \\
Rainy percentage & $3\%$ & $10\%$ & $27\%$ & $20\%$ & $0\%$ & $0\%$ & $6\%$ & $11\%$ & $10\%$ & $6\%$ & $9\%$ & $0\%$ \\
Unclassified percentage & $72\%$ & $44\%$ & $7\%$ & $59\%$ & $59\%$ & $26\%$ & $54\%$ & $54\%$ & $42\%$ & $30\%$ & $34\%$ & $19\%$ \\
\midrule
Winter percentage & $63\%$ & $41\%$ & $1\%$ & $47\%$ & $98\%$ & $10\%$ & $28\%$ & $47\%$ & $19\%$ & $47\%$ & $68\%$ & $0\%$ \\
Spring percentage & $37\%$ & $59\%$ & $99\%$ & $53\%$ & $2\%$ & $90\%$ & $72\%$ & $53\%$ & $81\%$ & $53\%$ & $32\%$ & $100\%$ \\
\bottomrule
\end{tabularx}
\caption{Summary of the OMB statistics for each vehicle in figure \ref{fig: vehicle specific omb histograms}. The uncertainty in the mean is at most $0.1^{\circ}$C for each vehicle OMB dataset. Also shown is the percentage of observations from a specific vehicle that appear in the weather-specific and seasonal datasets discussed in section \ref{sec: Statistical analysis of observation-minus-background departures}. Observations which do not appear in any of the weather-specific datasets are accounted for in the unclassified field on this table.}
\label{ta: vehicle OMB statistics summary}
\end{sidewaystable}
\end{center}

\section{Conclusion} \label{sec: conclusion}

In this work we have examined a novel low-precision vehicle-based observation dataset obtained from a Met Office proof-of-concept trial. An overview of the quality-control (QC) applied to the vehicle-based observations was given. The data that passed QC were examined and compared with UKV $1.5$m-air-temperature and surface-air-temperature model data, roadside weather information station (RWIS) data and hourly Met Office integrated data archive system (MIDAS) data. Using these datasets, we explored the characteristics of the vehicle-based observations that passed QC.

The QC procedure consisted of four tests which assessed different aspects of the vehicle-based observations. Reports that did not have an observation of air temperature or the necessary metadata to be tested were removed prior to the QC procedure. The climatological range test (CRT), stuck instrument test (SIT) and global positioning system (GPS) test were applied in parallel. Both the SIT and GPS test required vehicle identification, which may be unavailable in other crowdsourced and opportunistic datasets due to privacy concerns. While the majority of observations passed the CRT and SIT, a substantial number of observations were flagged by the GPS test due to unsuitable GPS update settings on the smartphone application and poor GPS signal. The observations that passed these three QC tests were put through a final sensor ventilation test (SVT) which flagged observations from vehicles driving below a predetermined sensor ventilation threshold speed. The SVT flagged a sizable amount of data relative to the amount tested. The final quality-controlled dataset (QC-dataset) consisted of $25.6\%$ of the observations obtained from the Met Office trial.

Using the QC-dataset, we investigated the uncertainty present in vehicle-based observations by analysing two observation-minus-background datasets. One dataset used UKV $1.5$m-air temperature as the background (aOMB) and the other dataset that used UKV surface-air temperature as the background (sOMB). Examining the OMB statistics of the QC-dataset we found that the vehicle-based observations of air temperature were on average greater than the UKV model data and agreed more with surface-air temperature than $1.5$m-air temperature. This is expected as the vehicle-based observations likely measure air temperature nearer to the surface than to a height of $1.5$m. However, there are several possible contributing factors to this result such as the quantization error, the difference in height between the vehicle-based observations and the UKV grid box, or heat from the vehicle engine contaminating the vehicle-based observations. We also found that the sOMB uncertainty is greater than the aOMB uncertainty for the QC-dataset. This is likely because the UKV surface-energy balance has a much stronger influence on surface-air temperature than $1.5$m-air temperature. 

To examine how the vehicle-based observation uncertainty changes with weather conditions, we grouped the OMB datasets by sunny, cloudy and rainy weather. For the sunny and cloudy OMB datasets, UKV surface-air temperature was on average greater than the vehicle-based observations with the magnitude of the sOMB bias greatest for the sunny dataset when the vertical air-temperature gradient near the surface is large. However, as shown in an illustrative time-series example, vehicle-based observations can be in greater agreement with $1.5$m-air temperature than surface-air temperature in sunny weather conditions. Possible explanations for this include the vehicle location (e.g., sea breeze effects) or the placement of the sensor used by the vehicle. For the rainy OMB dataset, when the vertical air-temperature gradient near the surface is small, we found that the vehicle-based observations were on average warmer than both UKV model fields. Inspecting the variability of the OMB datasets, we found that the sOMB variability is greatest in sunny weather conditions and smallest in rainy weather conditions. The aOMB variability was larger for the cloudy dataset than for the sunny dataset but still smallest for the rainy dataset. The large aOMB variability for the cloudy dataset may be due to changes in the sky-view factor caused by variable cloud cover or because it was the largest of the weather-specific datasets. These results strongly suggest that the uncertainty of vehicle-based observations of air temperature is weather-dependent. In particular, the uncertainty of vehicle-based observations will be largest in sunny weather conditions and smallest in rainy weather conditions.

To determine the effect of different vehicles on vehicle-based observations, the OMB datasets were grouped by vehicle. Due to the large number of observations with unclassified weather conditions and the different proportions of weather conditions experienced by vehicles, we were unable to distinguish between meteorological and vehicle effects on the vehicle-specific OMB distributions effectively. However, we note that vehicles with similar proportions of seasonal data may exhibit similar OMB variability. In order to determine the influence of the vehicle on the OMB statistics, further investigations into vehicle-based observations using a larger dataset must be conducted.


Vehicle-based observations are a potentially abundant source of low-cost, high-resolution meteorological information. However, there are several issues regarding data collection and privacy in addition to the uncertainty characteristics which must be addressed before they may be utilised. We first note that an increase in precision of the vehicle-based observations will allow improved QC, understanding of the characteristics of the data, and value for NWP. For QC, an alternative to vehicle identification must be used such that privacy concerns are mitigated. For assimilation, the uncertainty and bias associated with each observation must be sufficiently evaluated. This will require further trials to assess the effect of local meteorological conditions and the vehicle sensing instrument. Despite these issues, vehicle-based observations are a promising opportunistic dataset for convection-permitting data assimilation.

\section*{Acknowledgements}

This work was supported in part by a University of Reading PhD studentship and the EPSRC DARE project  EP/P002331/1. We would like to thank Katie O'Boyle and Malcolm Kitchen for conducting the Met Office trial which provided the vehicle-based observations and their helpful feedback on this work. We would also like to thank Ellie Creed for her assistance as well as additional members of the observations and road forecasting teams at the Met Office for the useful discussions. We thank Diego de Pablos for his work on vehicle-based observations obtained from a previous Met Office trial which provided a useful starting point for our work. The expertise of Martin Best and Peter Clark has been much appreciated for our own understanding and the discussion given in this paper. We would finally like to thank Siegfried Mercelis, Sylvain Watelet and Wim Casteels for providing us with additional explanation on their work using vehicle-based observations.

\section*{Data statement}

The Met Office trial dataset is not available in order to protect the privacy of the volunteer participants. We would like to thank Highways England for granting permission to use the roadside weather information station data \citep{rwis2018he}. The MIDAS and UKV datasets used in this study are available from the British Atmospheric Data Centre \citep{met2012met,met2016ukv}.

\section*{Disclosure}

No potential conflict of interest was reported by the authors.

\bibliography{bell_et_al_arxiv}

\end{document}